\documentclass[11pt,a4paper]{article}
\usepackage{lineno}
\usepackage{xr-hyper}
\usepackage{longtable}
\usepackage[utf8]{inputenc}
\usepackage{amsmath}
\usepackage{amsthm}
\usepackage[bitstream-charter]{mathdesign}
\usepackage{siunitx}
\usepackage{adjustbox}

\usepackage{xcolor}
\usepackage{orcidlink}
\usepackage{fontawesome}
\usepackage{tikz}
\usepackage{graphicx}
\usepackage[margin=1in]{geometry}
\usepackage{float}
\usepackage{verbatimbox}
\usepackage{booktabs}
\usepackage{algorithm}
\usepackage{algorithmic}
\usepackage{authblk}
\usepackage{lipsum}
\newcommand{\dis }{\displaystyle}
\usepackage{caption}
\usepackage{subcaption} 
\usepackage[T1]{fontenc}
\theoremstyle{definition}

\renewcommand{\arraystretch}{1.5}

\makeatletter
\def\thm@space@setup{\thm@preskip=1.2\parskip \thm@postskip=0pt}
\makeatother
\usepackage{tikz}
\usetikzlibrary{arrows}
\usepackage[toc,page]{appendix}
\usepackage{xurl} 
\usepackage[authoryear, round]{natbib}

\usepackage{hyperref}\hypersetup{
	colorlinks=true,
	citecolor=blue,
	linkcolor=red,
	filecolor=blue,
	urlcolor=blue,
}

\begin{document}
	\pagenumbering{gobble}
	
	\section*{Graphical Abstract}
	\vspace{0.8cm}
	\begin{figure}[H]
		\centering
		\includegraphics[width=\textwidth]{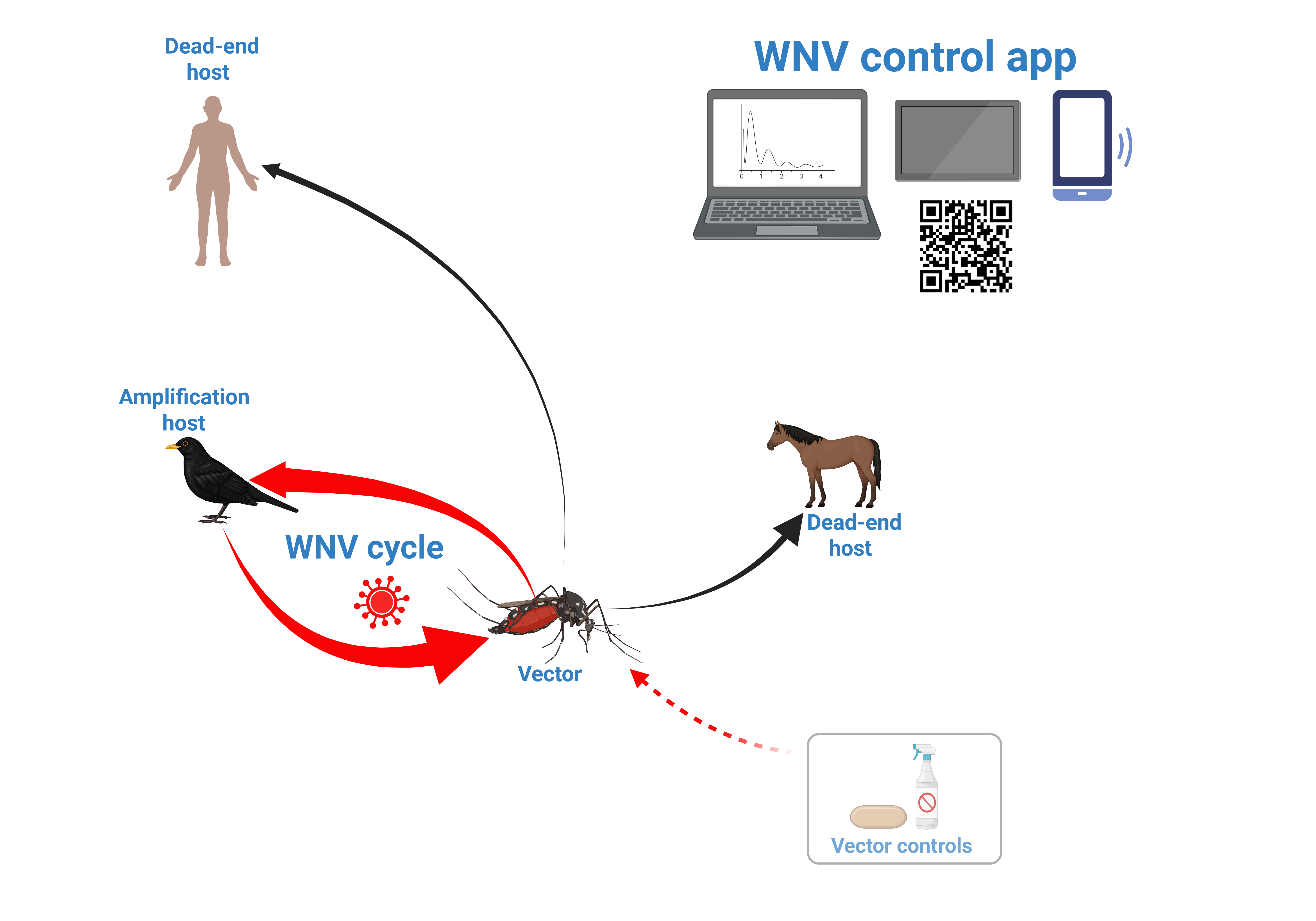}
		\label{f0}
	\end{figure}
	\vspace{0.2cm}
	\begin{center}
	\end{center}
	
	\newpage
	\section*{Highlights}
	
	\vspace{1em}
	\begin{itemize}
		\item A temperature-driven mechanistic model for WNV transmission is studied.
		\item Early-season mosquito controls most effectively reduce the August peak in birds.
		\item Behavioural responses triggered by bird surveillance reduce human spillover risk.
		\item Equid vaccination lowers infections, especially when combined with mosquito control.
		\item A Shiny decision-support tool explores WNV control under future climate scenarios.
	\end{itemize}
	
	\newpage
	
	\clearpage 
	\pagenumbering{arabic} 
	\setcounter{page}{1}

	\title{Modelling the control of West Nile virus using mosquito reduction methods, vaccination of equids, and human behavioural adaptation to the usage of personal protective equipment}

	\author{
		Pride Duve\,\orcidlink{0000-0002-0563-7292}\thanks{Corresponding author: \texttt{pride.duve@bnitm.de}}, \;Felix Gregor Sauer, and Renke Lühken \\
		\small	Bernhard Nocht Institute for Tropical Medicine, Hamburg, Germany}
	
	\date{}
	\maketitle

	\begin{abstract}
	West Nile virus (WNV) is a mosquito-borne virus in the genus \textit{Flavivirus} that circulates between mosquitoes and birds, whereas humans, equids, and other mammals are dead-end hosts. Since its emergence in Germany in 2018, the virus has spread across the country, emphasising the need for effective intervention strategies. However, it remains unclear how different strategies should be combined and timed to effectively reduce WNV transmission under temperature-driven dynamics. In this study, we develop a temperature-dependent, process-based model to evaluate the effectiveness of WNV control strategies, such as mosquito reduction methods, equid vaccination, and the use of personal protective equipment (PPE). Human behavioural responses to infection risk are incorporated through imitation dynamics that capture how individuals adopt PPE based on perceived infection risk and social influence. An optimal control problem has been formulated and studied to determine the seasonal timing of mosquito controls under temperature forcing. Results suggest that mosquito control efforts initiated in early spring and intensified in early May, may reduce the August peak in the infectious bird population. Moreover, a combined scenario of mosquito control methods, human PPE adoption, and equid vaccination could be the best strategy among dead-end hosts. The analysis of various combinations of constant controls is available as an interactive application, allowing users to explore intervention strategies under different temperature projections corresponding to the low-mitigation (SSP126), intermediate (SSP245), and high-emission (SSP585) scenarios.		
		
		\vspace{2em}
			\textbf{Keywords:} West Nile virus; temperature; mosquito control; behavioural dynamics; equid vaccination; Control app
	\end{abstract}

	\section{Introduction}
	
	The circulation of West Nile virus (WNV) in Europe has been observed for several decades \citep{Sambri2013}. Today, many countries in Southern and Southeastern Europe have confirmed cases in birds, equids, humans, and other mammals, while several other countries remain at risk of WNV  \citep{Bakonyi2020}. Climate warming has a significant role in the establishment and spread of WNV globally, including Europe \citep{Erazo2024}, posing a substantial public health and economic challenge \citep{Paz2015, Watts2021, zzziegler2019, Ziegler2019, Ziegler2020}. The enzootic cycle involves different \textit{Culex} species and birds, with a risk of spillover to humans, equids, and other mammals \citep{Vogels2016, Vogels2017}.
	
	Mammals are dead-end hosts as they do not develop viremia high enough to infect mosquitoes, while birds are considered the amplifying host \citep{Nemeth2007}. Most cases in humans are asymptomatic, but mild symptoms may include fever, vomiting, or skin rash. In contrast, severe symptoms may include fever, headache, or confusion, among others, and in some cases, the infection can be fatal \citep{Petersen2013, Sambri2013}. Symptoms in equids include ataxia, weakness, depression, dysphagia, and frequent stumbling \citep{Salazar2004, Schuler2004}. WNV symptoms in birds vary by species, but some species are at risk of developing ataxia, abnormal head posture, and rapid death \citep{Phalen2004}.
	
\begin{figure}[H]
		\centering
		\begin{subfigure}{0.4\textwidth}
			\centering
			\includegraphics[width=\linewidth]{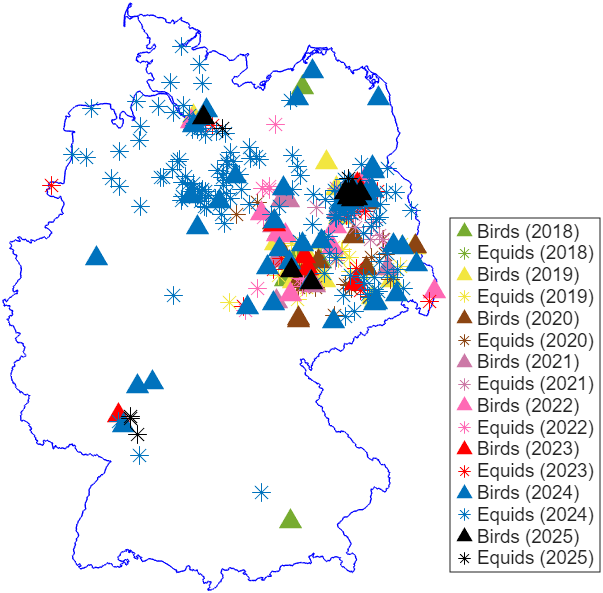}
			\label{f1}
		\end{subfigure}
		\hfill
		\begin{subfigure}{0.55\textwidth}
			\centering
			\includegraphics[width=\linewidth]{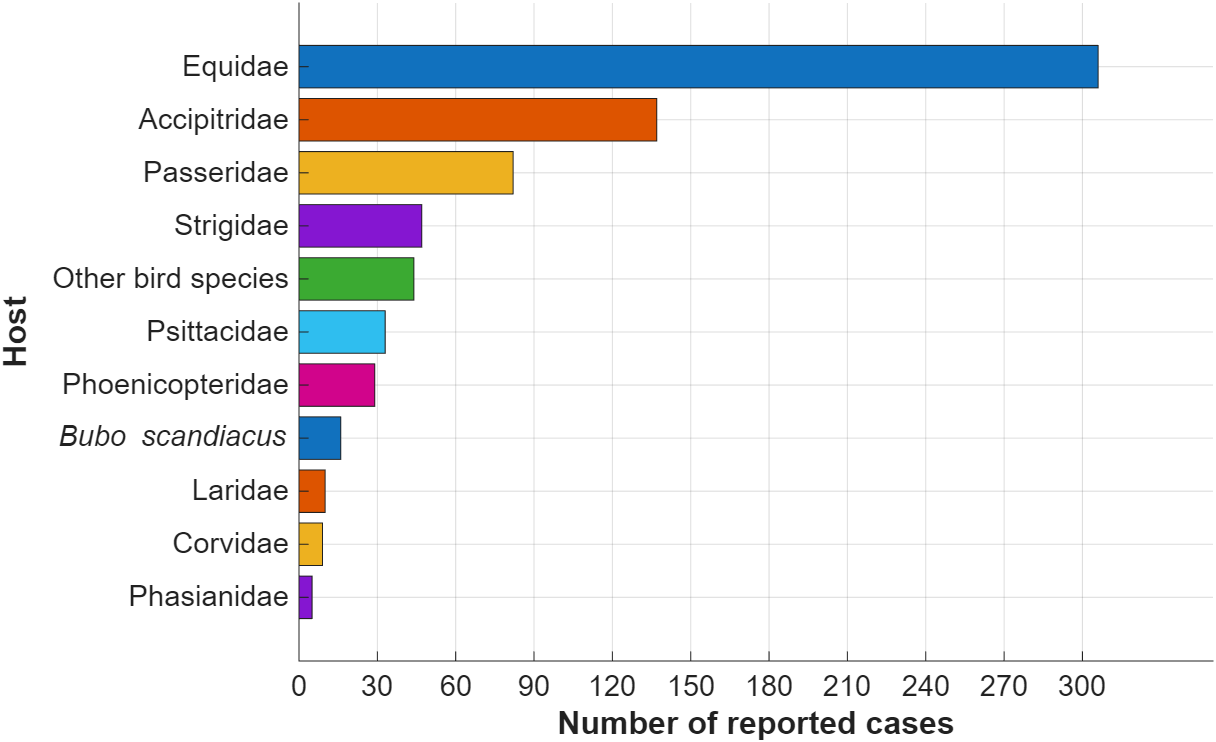}
			\label{f2}
		\end{subfigure}
		\caption{\small Observed animal WNV cases in Germany from August 2018 to December 2025 (left) and WNV cases for the different species (right). Data obtained from \citet{klji} for the period 2018-2025.}
		\label{fb}
	\end{figure}

	In Germany, the first case of WNV was reported in August 2018, in a bird from Eastern Germany \citep{zzziegler2019, Ziegler2019, Ziegler2020}. Over the following years, annual circulation was observed, particularly in central-eastern Germany (Fig. \ref{fb}). To date, there is no specific cure for WNV, and treatment only relies on managing symptoms \citep{Colpitts2012}. Licensed WNV vaccines are only available for equids \citep{Sambri2013}. Thus, preventive measures for humans rely mainly on recommendations to use personal protective measures that reduce contact between mosquitoes and humans \citep{Petersen2013, wst}, and, most importantly, on measures to reduce mosquito populations \citep{Bellini2014}. The European Centre for Disease Control and Prevention further stresses that people should consider using mosquito bed nets, sleeping in screened or air-conditioned rooms, wearing clothing that covers most of the body, or using mosquito repellents \citep{wst}. Thereby, several mosquito control methods are available, e.g., breeding-site removal, larvicides, adulticides, and biological control methods such as the introduction of natural enemies \citep{Cailly2012, Ferede2018, Takken2009}. 
	Several process-based models have been developed to assess various control measures and their timing for WNV. Early modelling studies focused on describing the interaction between mosquito vectors and avian hosts to understand the ecological drivers of WNV transmission \citep{Bhowmick2023, Laperriere2011, Mbaoma2024}. These models provided valuable insights into how environmental factors and host-vector interactions influence the spread of WNV. Subsequent studies then extended these frameworks by incorporating additional host populations and intervention strategies, such as mosquito control and vaccination. \citet{BOWMAN2005} found that adulticiding is more effective than personal protection for controlling the spread of human WNV. According to \citet{Thomas2009}, mosquito spraying during autumn is more effective than in summer. \citet{bhowmick2024host} demonstrated that ultra-low volume spraying only temporarily reduces the reproductive number. The combination of bird immunisation and mosquito control approaches is the most cost-effective strategy against WNV \citep{Malik2018}. \citet{Blayneh2010} concluded that mosquito reduction should be prioritised over personal protection measures, and \citet{abdelrazec2015dynamics} highlighted larviciding as the most effective ongoing strategy while emphasising the importance of seasonal timing. 
	
	Although these models have improved our understanding of WNV dynamics, many assume constant intervention strategies and do not explicitly consider how environmental factors, particularly temperature, influence the timing and effectiveness of mosquito control measures. Additionally, human behavioural responses, such as the adoption of personal protective equipment, are rarely incorporated into WNV transmission models despite their potential importance in reducing spillover infections. These limitations highlight the need for modelling frameworks that integrate environmental drivers, behavioural responses, and multiple intervention strategies to better inform WNV control.
	
	An interdisciplinary perspective, such as the One Health approach, is particularly important for zoonotic pathogens like WNV, as it integrates knowledge from ecology, animal health, and human health to design and evaluate control strategies \citep{Cendejas2024}. In this study, we analyse the effectiveness of several intervention strategies within a system of ordinary differential equations (ODEs), including larviciding, adult mosquito control, and PPE adoption. In addition, we incorporate the physical removal of mosquito breeding sites and the vaccination of equids, two interventions that have not been explicitly considered together in previous modelling studies. The model is driven by real-world temperature data for Berlin, allowing environmental conditions to influence mosquito dynamics and the effectiveness of control measures. Furthermore, we formulate and solve an optimisation problem to determine the timing of mosquito control strategies, driven by temperature. Finally, we explore constant intervention levels ranging from 0 to 90\% and evaluate different combinations of controls. These scenarios are implemented in our interactive, open-access web-based application, which enables users, including non-specialists, to explore the effects of different intervention strategies and their combinations without requiring expertise in programming or mathematical modelling.
	\section{Model formulation}\label{f}
	Our model consists of a system of first-order ODEs, extending the work of \citet{Laperriere2011}, \citet{Mbaoma2024}, and \citet{rubel2008}. We here focus on capturing control methods against WNV. Temperature-dependent parameters of the model are fitted from various datasets in \citet{Laperriere2011} and \citet{rubel2008}, and thus, for a more detailed derivation of the baseline WNV model and the thermal parameters, we refer the reader to the studies by \citet{Laperriere2011} and \citet{rubel2008}.
	
	\subsection{Mosquito population}
	
	In this study, the mosquito population is divided into four classes: mosquito larvae $(L_M),$ susceptible adults $(S_M),$ exposed adults $(E_M),$ and infectious adults $(I_M).$ The larval population increases through oviposition by adults, assuming that all adult mosquitoes under study are females. This recruitment process is proportional to the total adult mosquito population $ N_M=S_M+E_M+I_M$, thereby linking larval births to adult abundance. Overwintering dynamics are modelled via a hibernation rate $\delta_M$, which regulates the fraction of adult mosquitoes that lay eggs each year. Following \cite{rubel2008}, the fraction of non-diapausing adult mosquitoes is described by a photoperiod-dependent logistic function
	
	\begin{equation}\label{dM}
		\delta_M(D)=1 -\frac{1}{1 + 1775.7 \times \dis e^{\dis \left( 1.559 (D - 18.177) \right)}},
	\end{equation}
	
	where $D$ denotes the daytime length (in hours). The photoperiod depends on calendar day $d$ and geographic latitude $\phi$ through a standard astronomical relationship (see \citealp{rubel2008} for full derivation):
	$$D=7.639\arcsin\!\left[\tan(\epsilon)\tan(\phi)+\frac{0.0146}{\cos(\epsilon)\cos(\phi)}\right]+ 12,$$
	with solar declination 
	$$\epsilon=0.409\sin\!\left(\frac{2\pi(d-80)}{365}\right).$$
	
\begin{figure}[H]
		\centering
		\includegraphics[width=\textwidth]{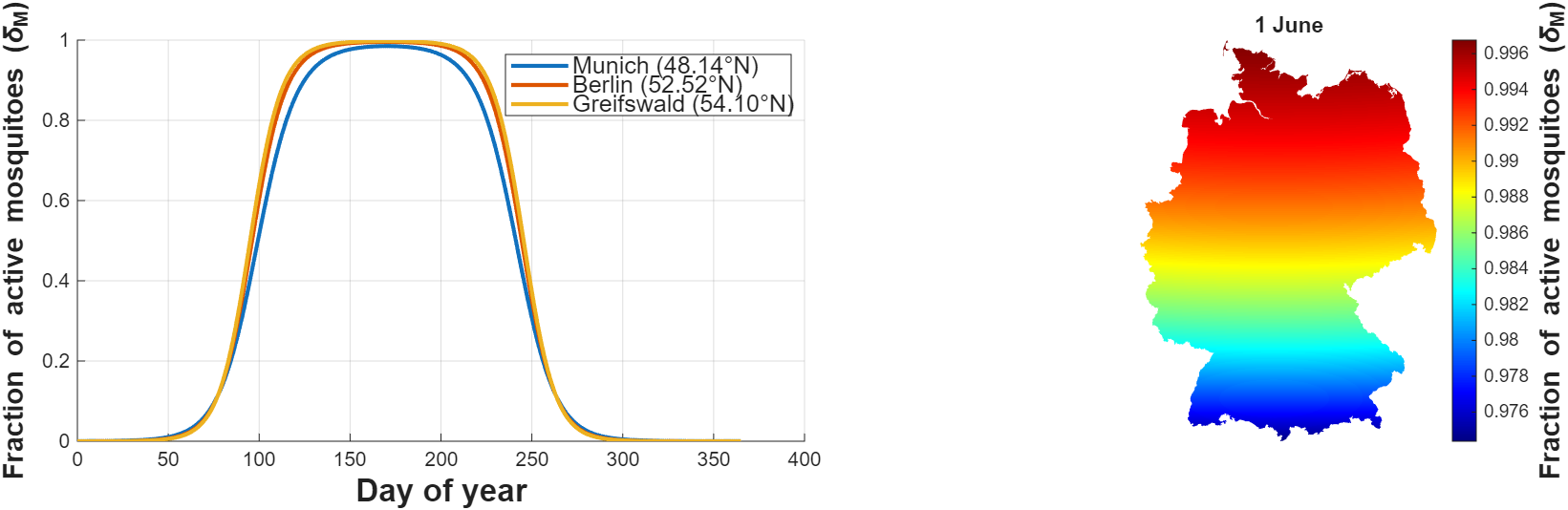}
		\caption{\small Fraction of active mosquitoes driven by photoperiod at selected locations (left) and for the entire country on 1 June (right).}
		\label{e0}
	\end{figure}

	The geographic latitude is fixed to Berlin ($\phi = 52.52^\circ \text{N}$), one of the WNV hotspots in Germany. Fig. (\ref{e0}) shows the spatio-temporal distribution of the function $\delta_M$ at selected locations to indicate its heterogeneity driven by the photoperiod over Germany.
	
	Although photoperiod primarily regulates diapause cues in temperate regions, daylight length alone is insufficient to determine the onset and intensity of transmission activity \citep{Eldridge1968, Field2022, Spielman1973}. Temperature acts as an additional environmental variable, modulating the general activity, development, survival, and feeding behaviour of mosquitoes \citep{Eldridge1968, rubel2008}. Moreover, viral replication within the mosquito vector is strongly temperature-dependent. Low temperatures prolong the extrinsic incubation period (EIP), potentially preventing mosquitoes from becoming infectious before they die. In contrast, higher temperatures accelerate viral replication and shorten the EIP, thereby increasing transmission potential \citep{Andreadis2014, Vollans2024}. 
	As a result, we here introduce temperature-dependent larval birth rate $(b_L(T)),$ mosquito development rate $(b_M(T))$, larvae mortality rate $(\mu_L(T)),$ adult mortality rate $(\mu_M(T)),$ biting rate $(k(T)),$ and the extrinsic incubation rate $(\gamma_M(T)),$ defined in Table (\ref{t2}). The temperature-dependent larval birth and death rates were fitted by \cite{rubel2008} and are given by:
	
	\begin{equation}\label{eq1}
		b_L(T)= \frac{0.7998}{1+1.231 e^{\dis -0.184(T-20)}},
	\end{equation}
	and
	\begin{equation}\label{eq3}
		\mu_L(T)= 0.0025T^2-0.094T+1.0257
	\end{equation}
	respectively. 
	Thus, the natural birth term of mosquito larvae governed by temperature and photoperiod is then given by:
	$$(1-\boldsymbol{u_1})b_L(T)\delta_M(D)\left[1-\frac{L_M}{K_M}\right]N_M,$$ 
	
	where $\dis \boldsymbol{u}_1\in[0,1]$ represents the level of implementation effort of breeding site removal. When $\boldsymbol{u}_1=0,$ then, the breeding site removal has no impact on larval production and mortality. In contrast, $\boldsymbol{u}_1=1$ corresponds to maximal effectiveness, meaning that full implementation may completely suppress larval recruitment and effectively increase its mortality. The term $\dis \left[1-\frac{L_M}{K_M}\right]$ is the density-dependent limiting factor in the breeding sites.
	
	As introduced earlier, the \cite{Ecdc2023} encourages the use of larvicides to increase the mortality of mosquito larvae; thus, we define a parameter $\boldsymbol{u}_2\in [0,1]$ to represent larviciding. Mosquito larvae that survive the larval stage develop into susceptible adults class $(S_M)$ at a temperature-dependent larval development rate:
	
	\begin{equation}\label{eq4}
		b_M(T)= \frac{b_L(T)}{10}.
	\end{equation}
	
	Susceptible mosquitoes get infected by biting infectious birds at a temperature-dependent mosquito-biting rate of 
	
	\begin{equation}\label{eq5}
		k(T)=\frac{0.344}{1+1.231 e^{\dis -0.184(T-20)}}
	\end{equation}
	
	and a probability that a successful bite of a susceptible mosquito on an infectious bird leads to a new mosquito infection $p_B=0.125.$ The force of infection on mosquitoes is thus given by
	
	\begin{equation}\label{eq6}
		\lambda_{BM}(T)=\frac{\delta_Mk(T)p_BI_B}{K_B},
	\end{equation}
	
	while the natural mortality rate of adult mosquitoes is given by the function
	
	\begin{equation}\label{eq7}
		\mu_M(T)= \frac{\mu_L(T)}{10}.
	\end{equation}
	
	For all adult mosquitoes, we consider the use of adult control measures denoted $\dis u_3$, which increase the adult mosquito mortality rate. Such measures can include adulticide sprays or traps that use mosquito lures to attract and capture adult mosquitoes, such as carbon dioxide traps, human-mimicking odours, or visual cues. Recently, \cite{Sauer2023} carried out a field study to identify overwintering sites for \textit{Cx. torrentium} and \textit{Cx. pipiens pipiens}. Their results indicated that \textit{Cx. torrentium} are mainly dominant in animal burrows, while \textit{Cx. pipiens pipiens} in human-made sites. Motivated by this overwintering behaviour of \textit{Culex} vectors, we include the elimination of adult-overwintering sites as part of adult control methods for \textit{Culex} mosquitoes. Although this control method has not yet been implemented in practice, it has been discussed in various studies, e.g., \citet{KOBAYASHI2012, Liu2016, Sauer2022}, given that overwintering mosquitoes are potentially easy to target and can be detected at high concentrations in their overwintering sites. After a successful interaction between susceptible mosquitoes and infectious birds, susceptible mosquitoes move to the exposed compartment $(E_M) $. The extrinsic incubation period in the exposed stage is given by $\dis \frac{1}{\gamma_M(T)},$ where 
	
	\begin{equation}\label{eq8}
		\gamma_M(T)=\begin{cases}
			\displaystyle 	0.0093T-0.1352\quad\text{if}\;& \displaystyle T>15,\\
			\displaystyle 	0 \quad & \text{otherwise},
		\end{cases}
	\end{equation}
	
	and from this stage, mosquitoes progress to the infectious class $(I_M).$ Temperature-dependent functions are summarised in Table (\ref{t2})and Fig (1) of the Supplementary material.

	\subsection{Bird population:}
	
	The primary bird species responsible for WNV amplification in Germany remains unknown, but it is suspected that multiple species may be involved \citep{Mbaoma2024, Michel2018, Offergeld2025}. Surveillance data from Germany indicate that birds from the Accipitridae and Passeridae families were the most affected species between the years 2018-2025 (Fig. \ref{fb} and \cite{Mbaoma2024, Ziegler2019}). The most frequently observed Accipitridae species during this period included the northern goshawk and other raptors, while the Eurasian blackbird (\textit{Turdus merula}) was among the Passeridae observed species \citep{Ziegler2019}. The blackbird is one of Germany's most common breeding birds, with approximately 10 million breeding pairs reported nationwide \citep{gerlach2019vogel}. Considering the high population of blackbirds and the frequent WNV infections observed in northern goshawks, these two species are used in this study as representative avian hosts.
	
	\cite{rubel2008} modelled blackbird birth rates by fitting a Gamma distribution to the observed frequency of blackbird nestlings, based on data from Poland. The resulting curve has the form 
	
	\begin{equation}\label{eq}
		b_{B_{b.b}}(d)= \frac{1.25}{10}\frac{(x/\beta)^{\alpha-1}\text{exp}(-x/\beta)}{\beta \Gamma (\alpha)}, \quad x,\alpha,\beta>0,
	\end{equation}
	
	which produces a right-skewed birth-rate function with a peak in nestling frequency around day 115.
	
	For northern goshawks, we used surveillance data from a citizen science project published by  \cite{alexandrecourtiol20204271624, MerlingdeChapa2020}. The data showed that urban goshawks breed earlier than their rural counterparts. Since our model is specific to Berlin, we extracted only Berlin urban data and fitted it to the Gamma distribution in Eq. (\ref{em}), with $x=d-d_0$. This resulted in fitted parameters $\alpha= 6.867, \beta= 2.935, d_0= 98,$ with a predicted peak in nestling frequency around day 115, the same day predicted by the function by \cite{rubel2008}. Here, $d_0$ marks the start of the breeding season. Based on this data, the annual per-capita recruitment rate was estimated as 1.5614 offspring per adult per year, leading to an estimated birth rate function of northern goshawks: 
	
	\begin{equation}\label{em}
		b_{B_{n.g}}(d)= 1.5614\frac{(x/\beta)^{\alpha-1}\text{exp}(-x/\beta)}{\beta \Gamma (\alpha)}, \quad x,\alpha,\beta>0.
	\end{equation}

	\begin{figure}[H]
		\centering
		\begin{subfigure}{0.48\textwidth}
			\centering
			\includegraphics[width=\linewidth]{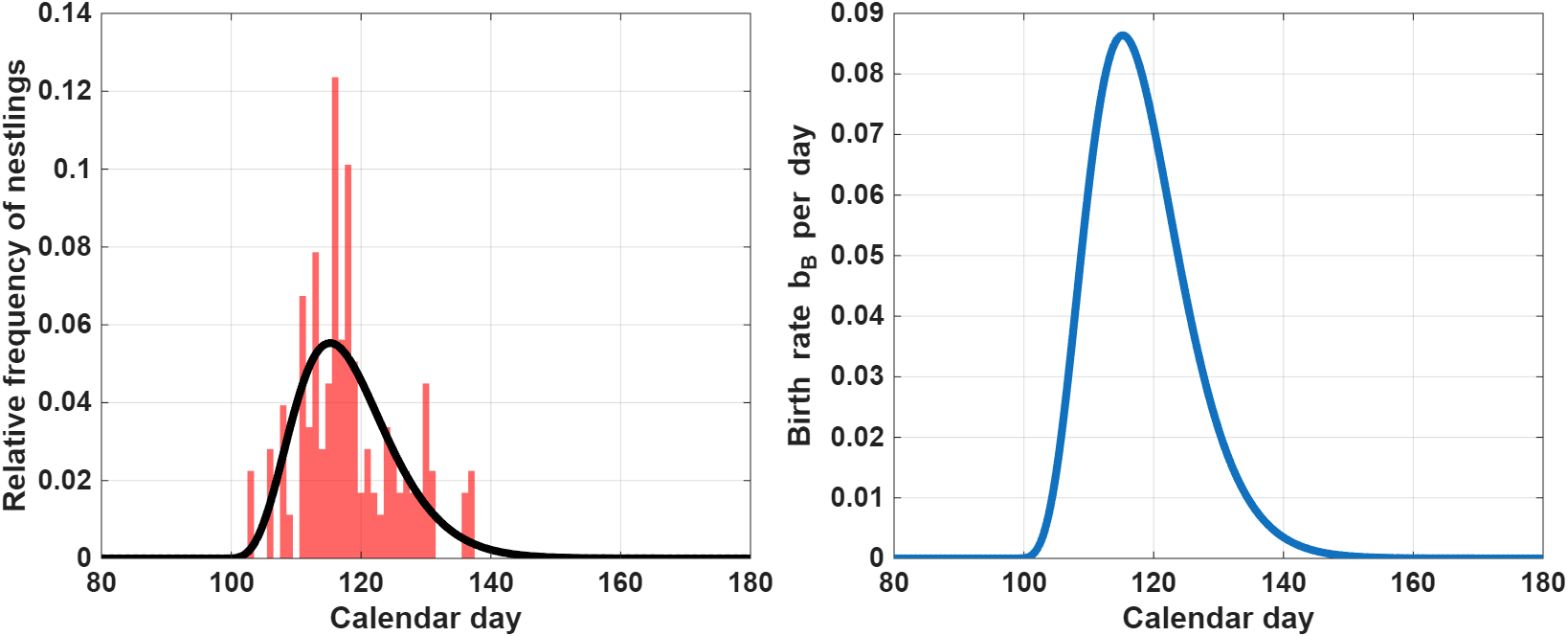}
			\label{w1}
		\end{subfigure}
		\hfill
		\begin{subfigure}{0.48\textwidth}
			\centering
			\includegraphics[width=\linewidth]{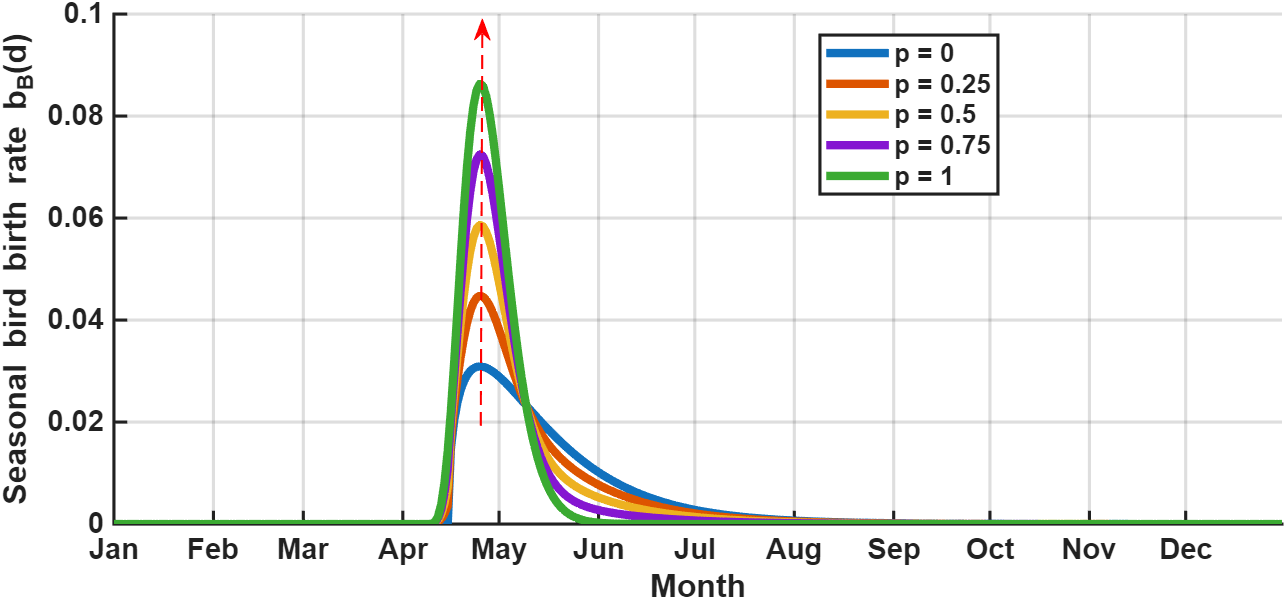}
			\label{w2}
		\end{subfigure}
		\caption{\small Observed relative frequency of the nestlings of northern goshawks \citep{alexandrecourtiol20204271624, MerlingdeChapa2020} with fitted Gamma distribution (left) and bird birth rate as a function of the calendar day (right).}
		\label{wb}
	\end{figure}
	
	The seasonal recruitment of susceptible birds is here modelled as a weighted mixture of the species-specific birth functions derived for northern goshawks and blackbirds. Such a formulation accounts for the fact that multiple bird species contribute to the pool of susceptible hosts involved in the transmission cycle. The combined recruitment function is therefore expressed as
	
	\begin{equation}\label{h}
		b_B(d) = p\, b_{B_{n.g}}(d) + (1-p)\, b_{B_{b.b}}(d),
	\end{equation}
	
	where $b_{B_{n.g}}(d)$ and $b_{B_{b.b}}(d)$ denote the seasonal birth functions for northern goshawks and blackbirds, respectively, and $d$ represents the day of the year. The parameter $0\leq p\leq 1$ determines the bird recruitment rate to the overall seasonal recruitment pattern, based on birth rates of northern goshawks and blackbirds. We examine the sensitivity of the seasonal birth-rate function $b_B(d)$ to the mixing parameter $p,$ illustrating how different values of $p$ influence the timing and magnitude of seasonal recruitment (Fig. \ref{wb}, right). The recruitment rate into the susceptible bird population is thus defined by
	
	$$b_{B}(d)\left[1-\frac{N_{B}}{K_{B}}\right]N_{B},$$
	
	while their natural mortality rate by $\mu_B=0.0005$.
	
	The population of birds is divided into susceptible $(S_B),$ exposed $(E_B),$ infectious $(I_B),$ recovered $(R_B),$ and dead birds $(D_B),$ with $N_B=S_B+E_B+I_B+R_B,$ and a force of infection between infectious mosquitoes and susceptible birds of
	
	\begin{equation}\label{eq10}
		\lambda_{MB}(T)=\frac{\delta_Mk(T)p_{M}\phi_BI_M}{K_{M}}.
	\end{equation}
	
	In Eq. (\ref{eq10}), $p_M=1.0$ is the probability that a successful bite by an infectious mosquito on a susceptible bird leads to a new bird infection, and $\dis \phi_B$ is the mosquito to bird ratio. The intrinsic incubation period of exposed birds is given by $\dis \frac{1}{\gamma_B},$ where $\dis \gamma_B=0.196,$ and from this stage, exposed birds progress to the infectious compartment. A proportion $\nu_B=0.655$ of infectious birds die due to the WNV, at a rate of $\alpha_B=0.867,$ while a proportion $(1-\nu_B)$ recover from the virus at a rate of $\alpha_B.$ Parameter definitions and sources are summarised in Table (\ref{t2}).
	
	\subsection{Human population:}
	
	\cite{Laperriere2011} and \cite{Mbaoma2024} modelled the human population by dividing it into susceptible $(S_H) $, exposed $(E_H) $, infected $(I_H) $, recovered $(R_H) $, and deceased $(D_H)$ compartments. In this study, we adopt these four compartments and integrate human behavioural adaptation to PPE use through game-theoretic and imitation-dynamics modelling.

	 \cite{Bauch2005} initially examined imitation dynamics to predict vaccination behaviour based on disease prevalence. We here use this approach to model how people may decide to start using PPE in response to WNV prevalence in birds. In such a formulation, PPE use is assumed to begin once an increase in WNV cases among birds is observed, as bird infections are typically detected before human cases \citep{Tamba2024}. Thus, the proportion of humans using PPE is here defined by $x,$ while $1-x$ are those who are unprotected at any time $t,$ with the rate of change of $x$ described by the equation 
	
	\begin{equation}\label{eq11}
		\frac{dx}{dt}= q x(1-x)\left[-r_p+r_imI_B\right].
	\end{equation}
	
	In this equation, $q$ is the imitation rate that measures how quickly people adopt PPE due to factors such as social influence, imitation, what they hear, and what they observe within their social circle. $r_p$ measures the perceived inconvenience, frustration, or annoyance due to PPE use, $r_i$ measures the perceived infection risk, and $m$ indicates how much disease prevalence in birds affects PPE adoption by susceptible humans. Notably, when $\dis r_imI_B> r_p,$ more people use PPE, whereas when $\dis r_imI_B< r_p,$ fewer people do. This enables us to model PPE adoption as a game in which players are susceptible humans who are undecided about adopting PPE, considering their potential impacts on people's daily socio-economic, cultural, and emotional livelihoods \citep{Bauch2005, Nkonzi2022}. The reduced version of Eq. (\ref{eq11}) is given by Eq. (\ref{eq12}) and the full derivation is shown in Section S3 of the Supplementary material. 
		
	\begin{equation}\label{eq12}
		\frac{dx}{dt}= \kappa x(1-x)\left[-1+\omega I_B\right].
	\end{equation}
	
 It follows that when $ x=0$, individuals do not use any PPE, whereas when $ x=1$, the entire susceptible population adopts PPE, representing the maximum level of protective behaviour in the model.

The population of susceptible humans in Berlin is obtained from \cite{WorldPopReview2026Berlin}, which reported an estimated population of approximately 3,685,265 in 2024. This office further reported 33,749 births, and 37,686 in the Berlin-Brandenburg area \citep{StatBB} during the same year. Additionally, Berlin is one of the major cities with high migration populations \citep{Heider2020, StatBBMigration2024}, with a 2024 net migration of 27,107, as reported by \cite{StatBBMigration2024}. Based on this data, we estimate the crude per-capita daily recruitment rate $(b_H)$ of susceptible humans in Berlin as the sum of the crude birth rate and the crude net migration rates, yielding estimates of $b_H=0.0000452\;\text{per day}$ while the natural mortality rates of humans is calculated from the crude death rates of humans in Berlin yielding $\mu_H=0.0000280\; \text{per day}.$ Full calculations are presented in the Section S3 of the Supplementary material.

	The force of infection on humans is given by
	
	\begin{equation}\label{eq13}
		\lambda_{MH}=\frac{(1-x)\delta_Mk(T)p_M\phi_HI_M}{K_M}.
	\end{equation}
	
	Susceptible humans progress to the exposed class via the force of infection in Eq. (\ref{eq13}). In the exposed class, the intrinsic incubation period is given by $\dis \frac{1}{\gamma_H},$ where $\gamma_H=0.25.$ As dead-end hosts, humans do not infect mosquitoes, instead, infected humans either recover at a rate $\alpha_H=0.5$ or die at a rate $\nu_H=0.004$ due to WNV \citep{Laperriere2011, Mbaoma2024}. The mosquito to human ratio is assumed to be $\phi_H=0.3-3.$ 
	
	\subsection{Equid population:}
	
	The population of equids is divided into susceptible $(S_E),$ vaccinated $(V_E),$ exposed $(E_E),$  infected $(I_E),$ recovered $(R_E),$ and dead $(D_E).$ Official statistics from the Landesamt für Ländliche Entwicklung, Landwirtschaft und Flurneuordnung (LELF; the state of Brandenburg department responsible for rural development and agriculture) report that approximately 45,000 horses are kept in Berlin and Brandenburg combined \citep{LELF2026Pferde}. Given that there is no publicly available census of equids in Berlin, we use this statistic as the baseline regional susceptible equid population, while exposed, infected, recovered, and deceased are initialised to zero. Susceptible equids are recruited at a birth rate of $b_E=0.00016$ and die at a natural mortality rate of $\mu_E=0.00011$ adopted from \cite{Laperriere2011}. Mosquitoes can bite either unvaccinated or vaccinated equids, with a force of infection
	
	\begin{equation*}\label{eq14}
		\lambda_{ME}=\frac{\delta_Mk(T)p_M\phi_EI_E}{K_M}.
	\end{equation*}
	
	A successful bite from an infectious mosquito on susceptible equids moves them to the exposed compartment. The force of infection for vaccinated equids is controlled by the threshold $(1-c_E) $, where $c_E$ is the vaccine effectiveness. When $ c_E=0$, the vaccine is ineffective, and vaccinated equids progress to the exposed compartment just like unvaccinated ones. Conversely, when $c_E=1,$ the vaccine is 100\% effective, and no vaccinated equids will be infected with WNV. The population of exposed equids is then increased via a combined force infection: $\dis \lambda_{ME}S_E+(1-c_E)\lambda_{ME}V_E,$ and exposed equids can die naturally at a rate of $\mu_E.$ The intrinsic incubation period of equids is $\dis \frac{1}{\gamma_E},$ where $\gamma_E=0.05.$ Equids in all compartments may die naturally at a rate $\mu_E,$ while a proportion $\nu_E=0.04$ of infected equids may die due to WNV and progress to the $D_E$ compartment at a rate $\alpha_E=0.2.$ A proportion $1-\nu_E$ recover at a rate of $\alpha_E.$ The mosquito to bird ratio is assumed to be $\phi_E=3,$ while the vaccination rate ranges from 0 to 1, as we aim to investigate different levels of vaccination rates. The vaccine waning rate is estimated as 0.00190 per day (full derivation shown in Section S3 of the Supplementary material).

	\subsection{Model flowchart and system of equations}
	
\begin{figure}[H]
		\centering
		\includegraphics[width=\textwidth]{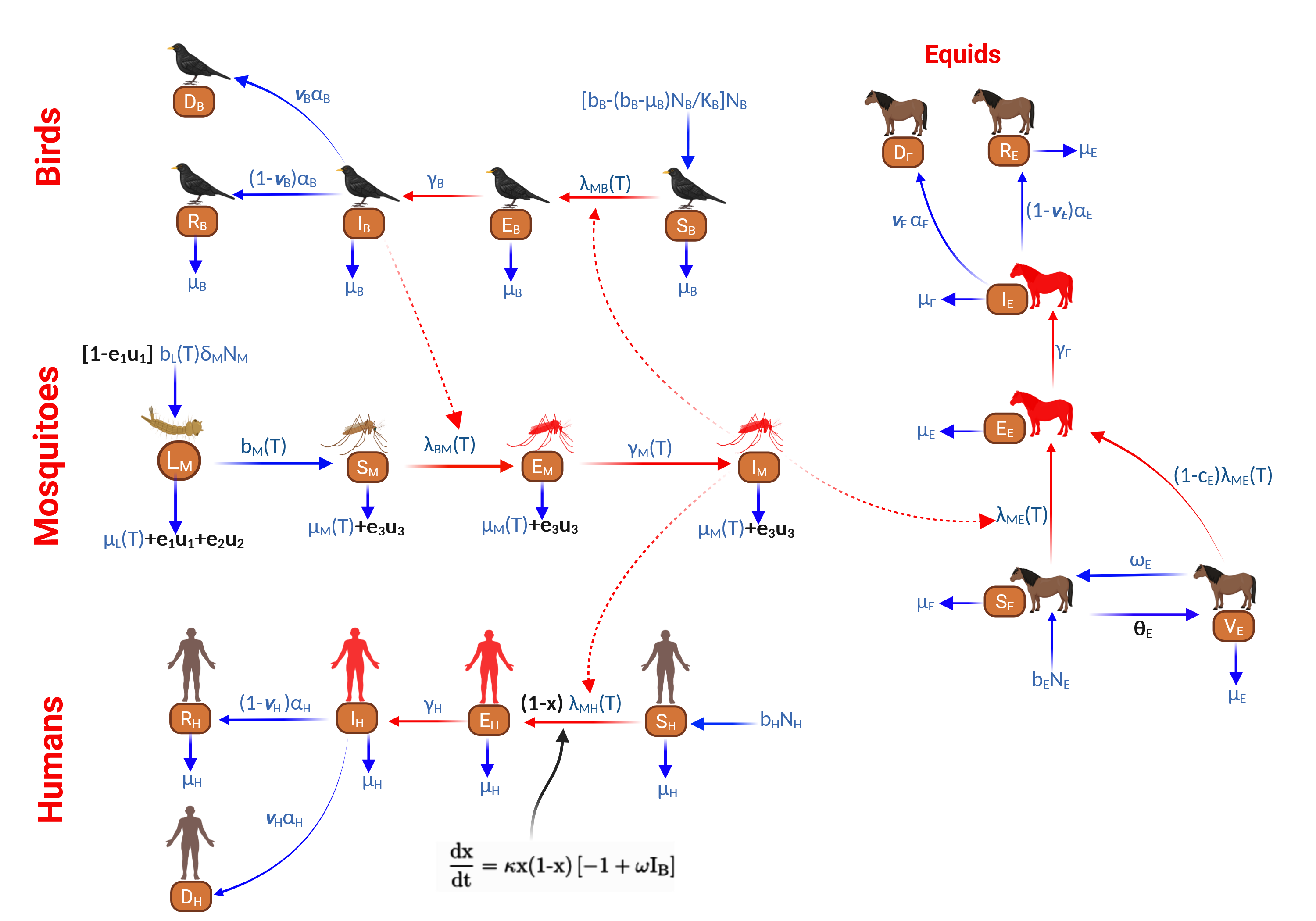}
		\caption{\small Flow chart diagram for the process-based model of WNV with control interventions in bold black colour. Red arrows and compartments indicate infected classes. Diagram created in BioRender with license Arbovirologie (2025) (https://BioRender.com/r9wek95). Content not licensed under the Creative Commons Attribution (CC BY) license.}
		\label{fg1}
	\end{figure}

	The flowchart diagram of our model is shown in Fig. (\ref{fg1}), and the complete set of equations for system (\ref{sys1a}) is given.
	\begin{equation}\label{sys1a}   
		\begin{split}	
			\text{Mosquitoes:} & \begin{cases}
				\displaystyle\frac{dL_M}{dt} & =\displaystyle\quad  \left[(1 -\boldsymbol{u_1})b_L(T)\delta_MN_M -(\mu_L+\boldsymbol{u_1+u_2})L_M\right]\left[1-\frac{L_M}{K_M}\right]-b_M(T)L_M,\\\\[-7pt]
				\displaystyle\frac{d S_M}{d t} & =\displaystyle\quad b_M(T)L_M -[\lambda_{B_M}(T)+\mu_M(T)+\boldsymbol{u_3}]S_M,\\\\[-7pt]			
				\displaystyle	\frac{d E_M}{d t} & =\quad \lambda_{B_M}(T)S_M-[\gamma_M(T)+\mu_M(T)+\boldsymbol{u_3}]E_M,\\\\[-7pt]
				\displaystyle	\frac{d I_M}{d t} & =\displaystyle \quad \gamma_M(T)E_M-[\mu_M(T)+\boldsymbol{u_3}]I_M,\end{cases}\\ \\[-5pt]
			\text{Birds:} & \begin{cases}
				\displaystyle	\frac{d S_{B}}{d t} & =\displaystyle \quad b_{B}\left[1 - \frac{N_{B}}{K_{B}}\right]N_{B}-\left[\lambda_{MB}(T)+\mu_{B}\right]S_{B},\\\\[-7pt]
				\displaystyle	\frac{d E_{B}}{d t} & =\displaystyle \quad \lambda_{MB}(T)S_{B}-[\gamma_{B}+\mu_{B}]E_{B},\\\\[-7pt]
				\displaystyle	\frac{d I_{B}}{d t} & =\displaystyle \quad \gamma_{B}E_{B}-[\alpha_{B}+\mu_{B}]I_{B},\\\\[-7pt]
				\displaystyle	\frac{d R_{B}}{d t} & =\displaystyle \quad (1-\nu_{B})\alpha_{B} I_{B}-\mu_{B}R_{B},\\\\[-7pt]
				\displaystyle	\frac{d D_{B}}{d t} & =\displaystyle\quad \alpha_{B}\nu_{B} I_{B},\\\end{cases}\\\\
			\text{Humans:} & \begin{cases}
				\displaystyle	\frac{d S_{H}}{d t} & =\displaystyle \quad b_HN_H-(1-\boldsymbol{x})\lambda_{MH}(T)S_{H}-\mu_HS_{H},\\\\[-7pt]
				\displaystyle	\frac{d E_{H}}{d t} & =\displaystyle \quad (1-\boldsymbol{x})\lambda_{MH}(T)S_{H}-(\mu_H+\gamma_{H})E_{H},\\\\[-7pt]
				\displaystyle	\frac{d I_{H}}{d t} & =\displaystyle \quad \gamma_{H}E_{H}-(\alpha_{H}+\mu_H)I_{H},\\\\[-7pt]
				\displaystyle	\frac{d R_{H}}{d t} & =\displaystyle \quad (1-\nu_{H})\alpha_{H} I_{H}-\mu_HR_H,\\\\[-7pt]
				\displaystyle	\frac{d D_{H}}{d t} & =\displaystyle\quad \nu_{H}\alpha_{H} I_{H},\\\\[-7pt]
				\displaystyle	\frac{d \boldsymbol{x}}{d t} & =\displaystyle \quad \kappa \boldsymbol{x}(1-\boldsymbol{x})\left[-1+\omega I_B\right],\end{cases}\\\\[-5pt]
			\text{Equids:} & \begin{cases}
				\displaystyle	\frac{d S_{E}}{d t} & =\displaystyle \quad b_EN_E+\omega_EV_E-\lambda_{ME}(T)S_E-(\mu_E+\boldsymbol{\theta_E})S_{E},\\\\[-7pt]
				\displaystyle	\frac{d V_{E}}{d t} & =\displaystyle \quad \boldsymbol{\theta_E}S_E-(\omega_E+\mu_E)V_E-(1-c_E)\lambda_{ME}(T)V_E,\\\\[-7pt]
				\displaystyle	\frac{d E_{E}}{d t} & =\displaystyle \quad \lambda_{ME}(T)S_{E}+(1-c_E)\lambda_{ME}(T)V_{E}-[\gamma_{E}+\mu_{E}]E_{E},\\\\[-7pt]
				\displaystyle	\frac{d I_{E}}{d t} & =\displaystyle \quad \gamma_{E}E_{E}-[\alpha_{E}+\mu_{E}]I_{E},\\\\[-7pt]
				\displaystyle	\frac{d R_{E}}{d t} & =\displaystyle \quad (1-\nu_{E})\alpha_{E} I_{E}-\mu_{E}R_{E},\\\\[-7pt]
				\displaystyle	\frac{d D_{E}}{d t} & =\displaystyle\quad \nu_{E}\alpha_{E} I_{E},\end{cases}
		\end{split}
	\end{equation}

	$\displaystyle \text{with}\;
	N_{M}=L_M+S_M+E_M+I_M,\;  N_{B}=S_{B}+E_{B}+I_{B}+R_{B},\; N_{H}=S_{H}+E_{H}+I_{H}+R_{H},\; N_{E}=S_{E}+V_E+E_{E}+I_{E}+R_{E}.$ Initial values used and a summary of the description of the state variables are shown in \textcolor{blue}{Table (S1)} of the Supplementary material. In contrast, parameter values are summarised in Table (\ref{t2}).

	\section{Methods}
	A global sensitivity analysis was conducted to evaluate the influence of control parameters on WNV transmission dynamics. Sensitivity indices were determined using the Partial Rank Correlation Coefficient (PRCC) method \citep{Blower1994, Marino2008}, which measures the strength of monotonic relationships between model inputs and outputs. Parameter values were sampled via Latin Hypercube Sampling, a stratified Monte Carlo technique, with 1000 simulations per run (Fig. \ref{sens}). 
	
	The model was then simulated for the period 2018-2025 using temperature data derived from the ERA5-Land Monthly Averaged Data from 1950 to Present for Berlin, obtained through the Copernicus Climate Data Store (CDS) \cite{copcds} (Figs. \ref{vc0}-\ref{vc2}). The data were first interpolated to daily resolution to provide temperature inputs for the model's temperature-dependent parameters, aligning with the solver's daily time step. Linear interpolation along the temporal dimension was performed using \texttt{xarray.interpolate\_na()} function in Python \citep{Hoyer2017}. Model solutions were scaled to the observed data, and Spearman's correlation coefficient was calculated between the monthly observed and simulated time series for each year to assess the similarity in seasonal epidemic dynamics.
	
	Control parameters were varied within their reasonable ranges to examine the impact of increased interventions on WNV dynamics. To explore different control scenarios, an interactive web-based Shiny application was developed to enable users to test different combinations of intervention strategies and their epidemiological outcomes in real time. Firstly, climate projections from the CMIP6 dataset in the CDS \citep{Copernicus2021CMIP6} were used to simulate the model for the years 2026-2046. Three shared socioeconomic pathway scenarios were considered, i.e., SSP1-2.6, SSP2-4.5, and SSP5-8.5, representing low, intermediate, and high greenhouse gas emission factors. These scenarios were included, to assess how different climate predictions may influence temperature-driven WNV dynamics, which has an impact on control intensity and timing. For each SSP scenario, controls can be implemented over user-defined time intervals and explored under each scenario. Our application was implemented using the \texttt{RShiny} package in \texttt{R} \citep{wes}, and is freely available as a free open-source tool at \href{https://zero-west-nile-virus.bnitm.de/}{\faExternalLink} or via the link: \url{https://zero-west-nile-virus.bnitm.de/}.
	
	In the app, users can specify the simulation period, and the start and end dates for implementing the controls (default: 1 January 2026 to 31 December 2046), and run the model by clicking the \textcolor{blue}{$\blacktriangleright$} button. The model is then solved dynamically, and simulations of infectious mosquitoes and birds, humans, and equids are immediately displayed. A 10\% threshold, defined as 10\% of the maximum simulated value for each compartment, is included to help users explore intervention combinations that lower epidemic curves below this level.
	
	To determine the necessary conditions for the optimal timing and effort level of mosquito controls under temperature forcing, an optimal control framework was formulated. For this, an objective functional (Eq. \ref{op}) was defined to minimise both the eco-epidemiological burden of WNV and the economic costs associated with implementing mosquito control over a 24-month period. This functional penalises high mosquito abundance (both larval and adults) and bird infections while also accounting for the costs of applying control strategies. The objective functional (Eq. \ref{op}) integrates three components: mosquito larvae, adult population, and infectious birds, together with quadratic costs associated with the three control measures: breeding-site removal $(u_1),$ larviciding $(u_2),$ and adult mosquito control $(u_3).$ Such a formulation helps us understand the necessary conditions under which mosquito control is most effective, based on temperature predictions, thereby minimising mosquito larvae, adult mosquitoes, and infectious bird populations at the minimum possible cost. The optimal control problem was solved using Pontryagin's Maximum Principle \citep{Pontryagin2018}, which introduces adjoint (costate) equations associated with the state variables. This framework was formulated by minimising the objective functional:
	
	\begin{equation}\label{op}
		J=\int_{0}^{T_F} A_1L_M+A_2N_M+A_3I_{B}+ \frac{1}{2}\left[c_1u_{1}^{2}(t)+ c_2u_{2}^{2}(t)+ c_3u_{3}^{2}(t)\right]\; dt,
	\end{equation}
	
	subject to the state system (\ref{sys1a}), without humans and equids, and the admissible control set
	
	$$
	\mathcal{U}=\left\{(u_1(t),u_2(t),u_3(t)) : 0 \leq u_i(t) \leq 0.5,\; t\in[0,T_F],\; i=1,2,3 \right\}.
	$$
	
	Here, $\dis A_1, A_2, A_3$ are normalised weight constants of the state variables, while $c_1, c_2, c_3$ are cost coefficients associated with the control variables, whose values are defined in Table (\ref{xc}). The functions $u_1(t), u_2(t), u_3(t)$ represent bounded Lebesgue measurable controls on the interval $[0, T_F]$, with \(T_F\) indicating the terminal time of controls. An upper bound of 0.5 was imposed on the control functions, assuming that the interventions may not be fully effective \citep{Blayneh2010}. The full derivation and proof are provided in the Supplementary material. The optimal control problem was solved numerically using the forward-backward sweep method \citep{Lenhart2007, rodrigues2014optimal}, where the state system (\ref{sys1a}) was solved forward in time and the co-state system (S8 in the Supplementary material) backward in time using a forward Euler scheme, together with the control characterisations (Eq. \ref{fg3}), until convergence. $\dis A_1, A_2, A_3$ were normalised using maximum values from the uncontrolled simulation, and their values chosen such that $\dis A_1> A_2; \; A_1, \; A_2<A_3,$ indicating that reducing infectious birds is more important than reducing mosquito abundance. Similarly, the cost coefficients $c_1,c_2,c_3$ are such that $c_3>c_1, c_2$, reflecting the very high costs associated with adult control methods compared to larviciding and breeding site removal (Table \ref{xc}).
	
	\begin{table}[H]
		\renewcommand{\arraystretch}{1.2}
		\small
		\centering
		\caption{\small Interpretation of the weight parameters and control costs in the objective functional, and their assumed values.}
		\begin{adjustbox}{width=1.0\textwidth}
			\begin{tabular}{cp{4.2cm}p{8cm}p{2cm}}
				\hline
				\textbf{Parameter} & \textbf{Description} & \textbf{Interpretation}&Value \\
				\hline
				$A_1$ & Weight associated with mosquito larvae ($L_M$) & 
				Represents the epidemiological cost associated with high mosquito larval abundance.& \text{0.0001/max(LM)}\\
				$A_2$ & Weight associated with adult mosquito population ($N_M$) &
				Represents the epidemiological cost of having a large number of adult mosquitoes that transmit WNV.& \text{0.0001/max(NM)}\\
				$A_3$ & Weight associated with infectious birds ($I_B$) &
				Represents the epidemiological cost of WNV infection in birds and the need to limit viral amplification.& \text{0.1/max(IB)}\\
				$c_1$ & Cost coefficient for breeding-site removal ($u_1$) &
				Represents the economic and operational costs associated with environmental management interventions such as habitat modification and removal of standing water.&50 \\
				$c_2$ & Cost coefficient for larviciding ($u_2$) &
				Represents the cost associated with the application of larval control measures aimed at killing mosquito larvae in breeding habitats.&50 \\
				$c_3$ & Cost coefficient for adult mosquito control ($u_3$) &
				Represents the economic and operational cost of adult mosquito control measures.&$10^4$ \\
				\bottomrule
			\end{tabular}
		\end{adjustbox}
		\label{xc}
	\end{table}

	\section{Results}
	
	Findings from the model fit, sensitivity analysis, behavioural adoption of PPE, vaccination of equids and the optimal control are presented here. 
	
	\subsection{Model fit}
	
	The observed data used in this study are very sparse because no systematic WNV surveillance was conducted in Germany during the study period. As a result, reported cases are sporadic and unevenly distributed over time. To enable a meaningful comparison between the model results and observations, both the simulated and observed data were aggregated into monthly totals to minimise daily fluctuations and highlight the overall seasonal patterns of WNV transmission \citep{Laperriere2011, rubel2008}. 
	
	\begin{figure}[H]
		\centering
		\begin{subfigure}{0.79\textwidth}
			\centering
			\includegraphics[width=\textwidth]{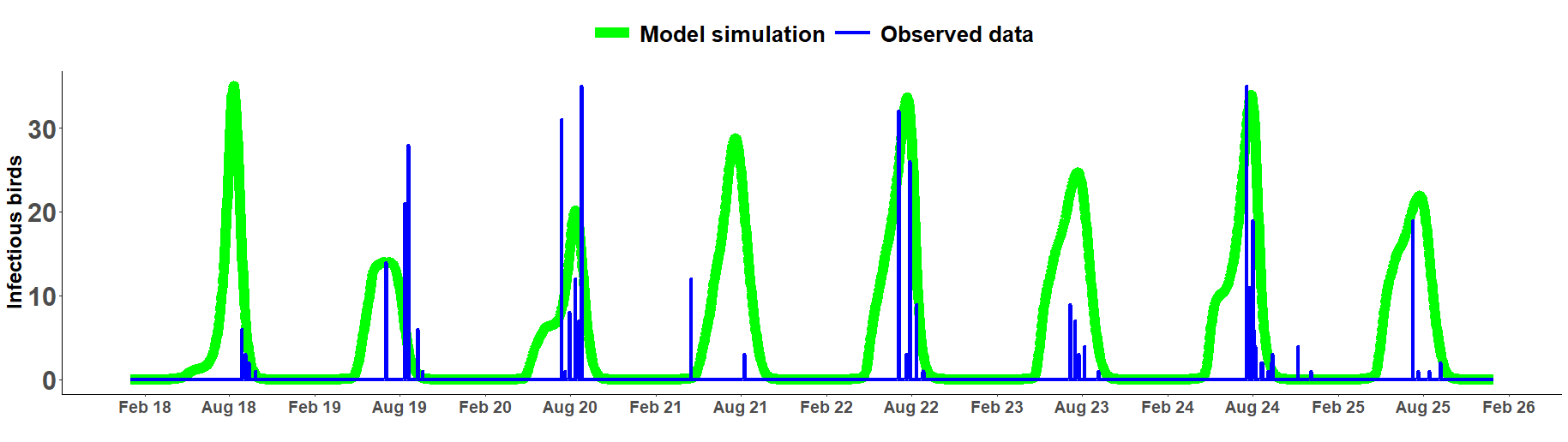}
			\caption{\small Infectious birds.} 
			\label{vc0}
		\end{subfigure}
		\hfill
		\begin{subfigure}{0.79\textwidth}
			\centering
			\includegraphics[width=\textwidth]{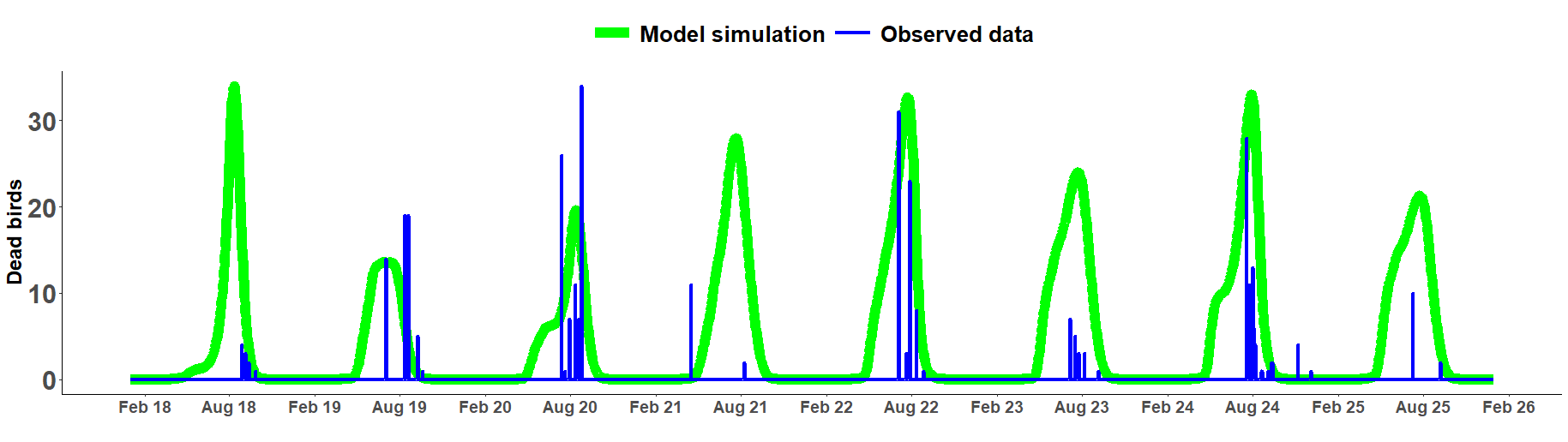}
			\caption{\small Dead birds.} 
			\label{vc1}
		\end{subfigure}
		\hfill
		\begin{subfigure}{0.79\textwidth}
			\hfill
			\includegraphics[width=\textwidth]{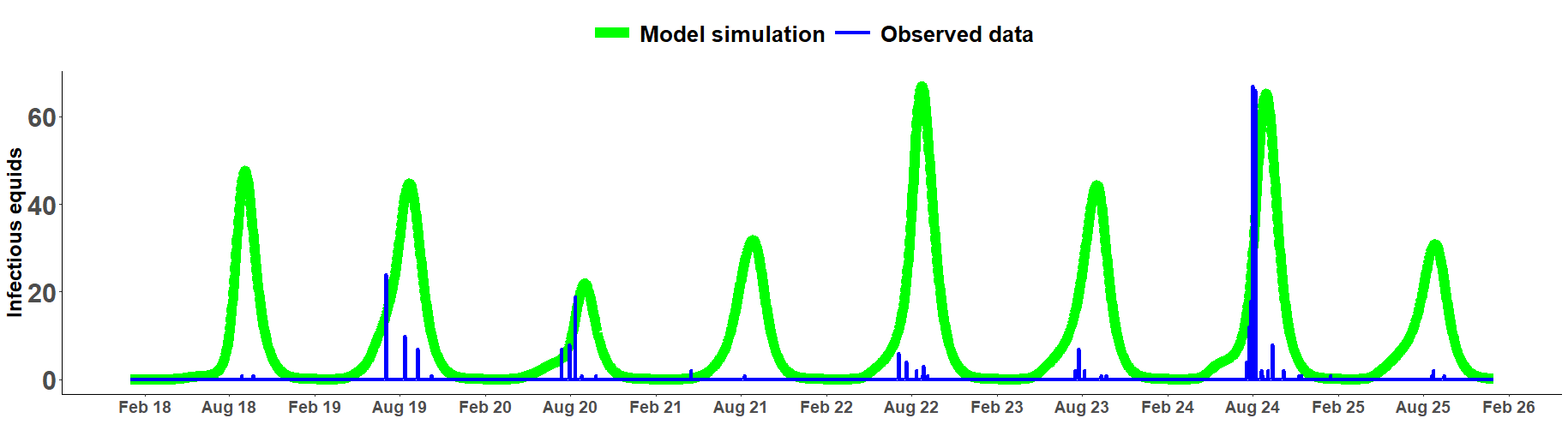}
			\caption{\small Infected equids.} 
			\label{vc2}
		\end{subfigure}
		\caption{\small Temporal dynamics of WNV compared to the observed cases for the period 2018-2025.}
		\label{vc}
	\end{figure}

	\begin{figure}[H]
		\centering
		\begin{subfigure}{0.79\textwidth}
			\centering
			\includegraphics[width=\textwidth]{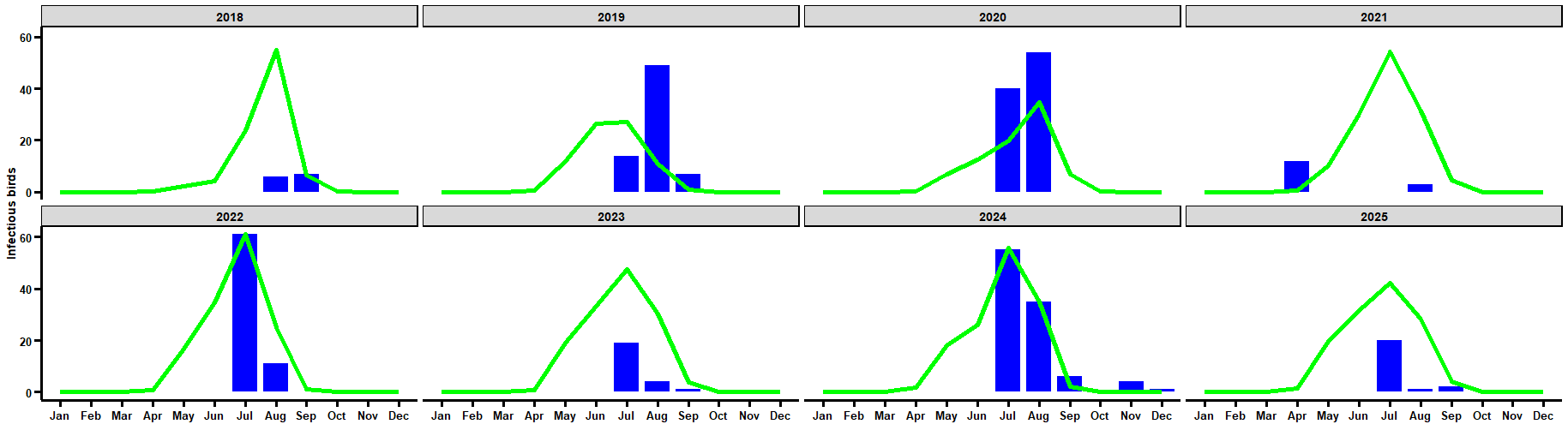}
			\caption{\small Infectious birds $(\rho=0.53).$} 
			\label{t0}
		\end{subfigure}
		\hfill
		\begin{subfigure}{0.79\textwidth}
			\centering
			\includegraphics[width=\textwidth]{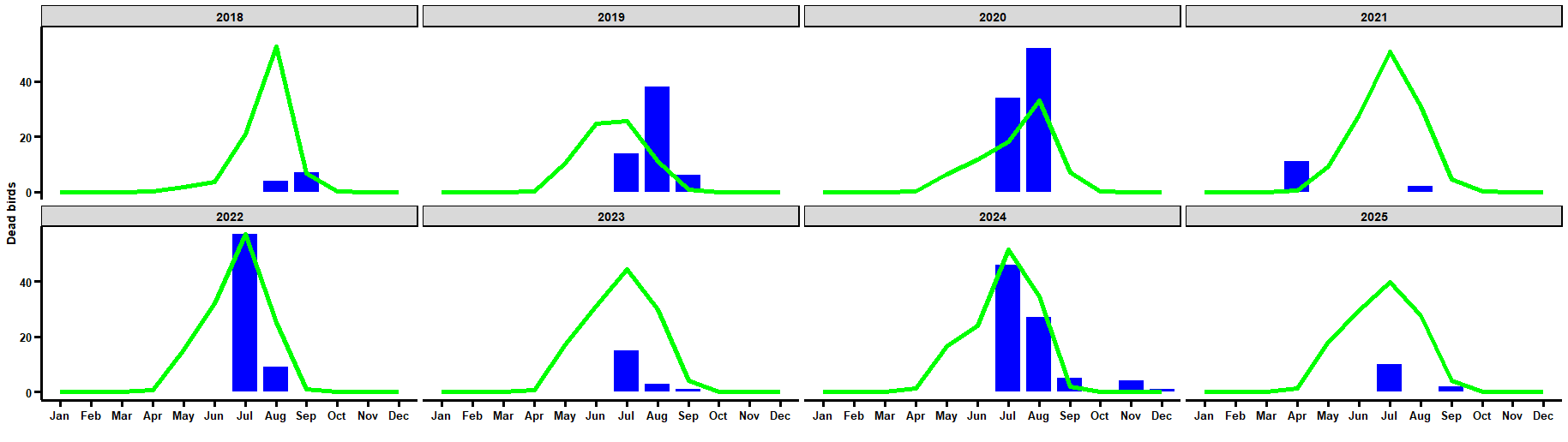}
			\caption{\small Dead birds $(\rho=0.52).$} 
			\label{t1}
		\end{subfigure}
		\hfill
		\begin{subfigure}{0.79\textwidth}
			\hfill
			\includegraphics[width=\textwidth]{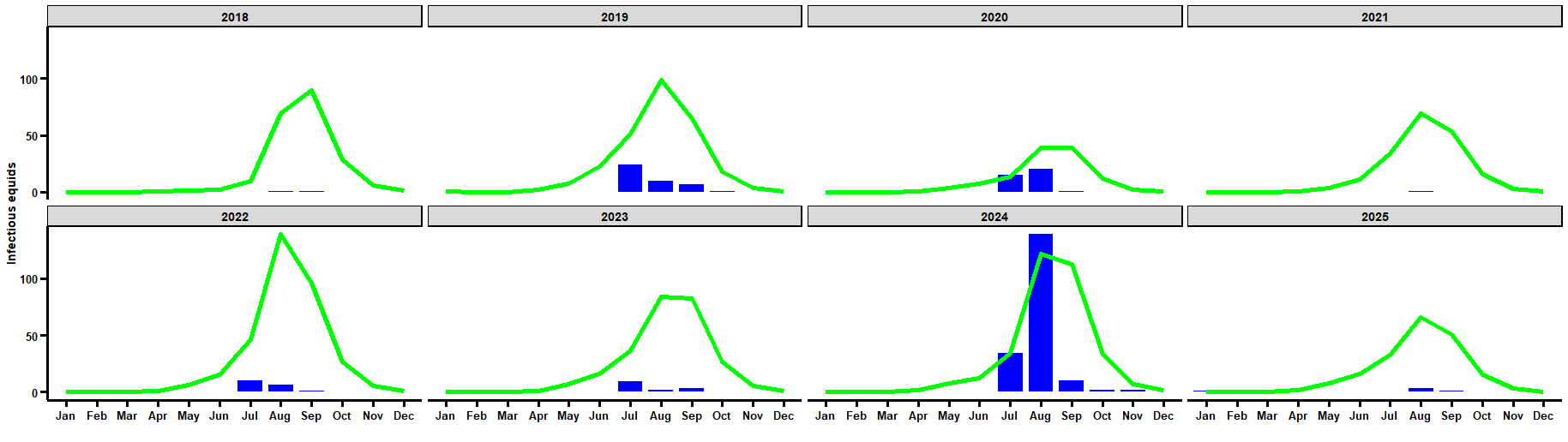}
			\caption{\small Infected equids $(\rho=0.63).$} 
			\label{tp2}
		\end{subfigure}
		\caption{\small Temporal comparison of observed and simulated WNV infections in birds (2018-2025). Blue bars show monthly observed cases, while the green line represents the corresponding monthly model simulations.}
		\label{v}
	\end{figure}	
	\subsection{Sensitivity analysis of the control variables}\label{g}
	Sensitivity analysis was performed to assess the contribution of each control parameter to the host populations. Negative PRCC values indicate that increasing the control effort is associated with a decrease in the corresponding infectious/infected host population. Conversely, positive PRCC values indicate that increasing the control effort increases the infected host population \citep{Marino2008}.

	\begin{figure}[H]
		\centering
		\begin{subfigure}{0.9\textwidth}
			\centering
			\includegraphics[width=\textwidth]{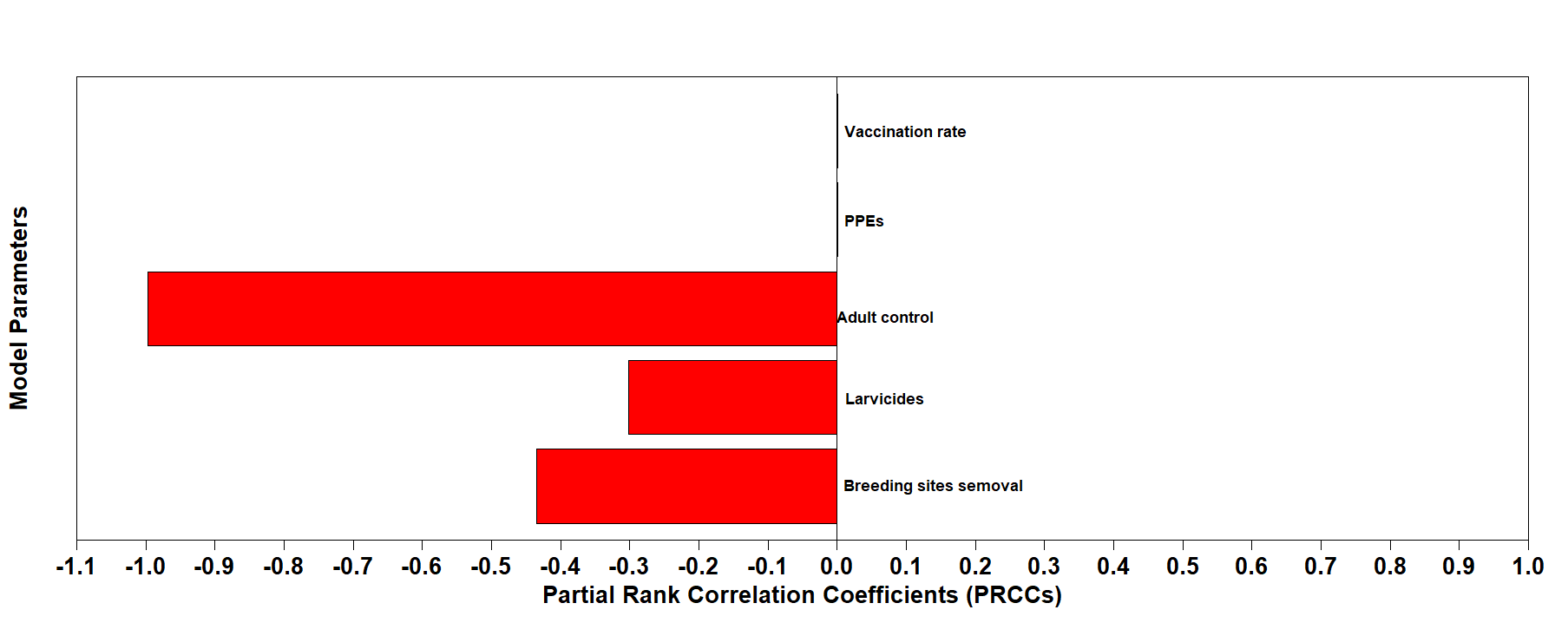}
			\caption{\small PRCC values for infectious birds.} 
			\label{s0}
		\end{subfigure}
		\hfill
		\begin{subfigure}{0.9\textwidth}
			\centering
			\includegraphics[width=\textwidth]{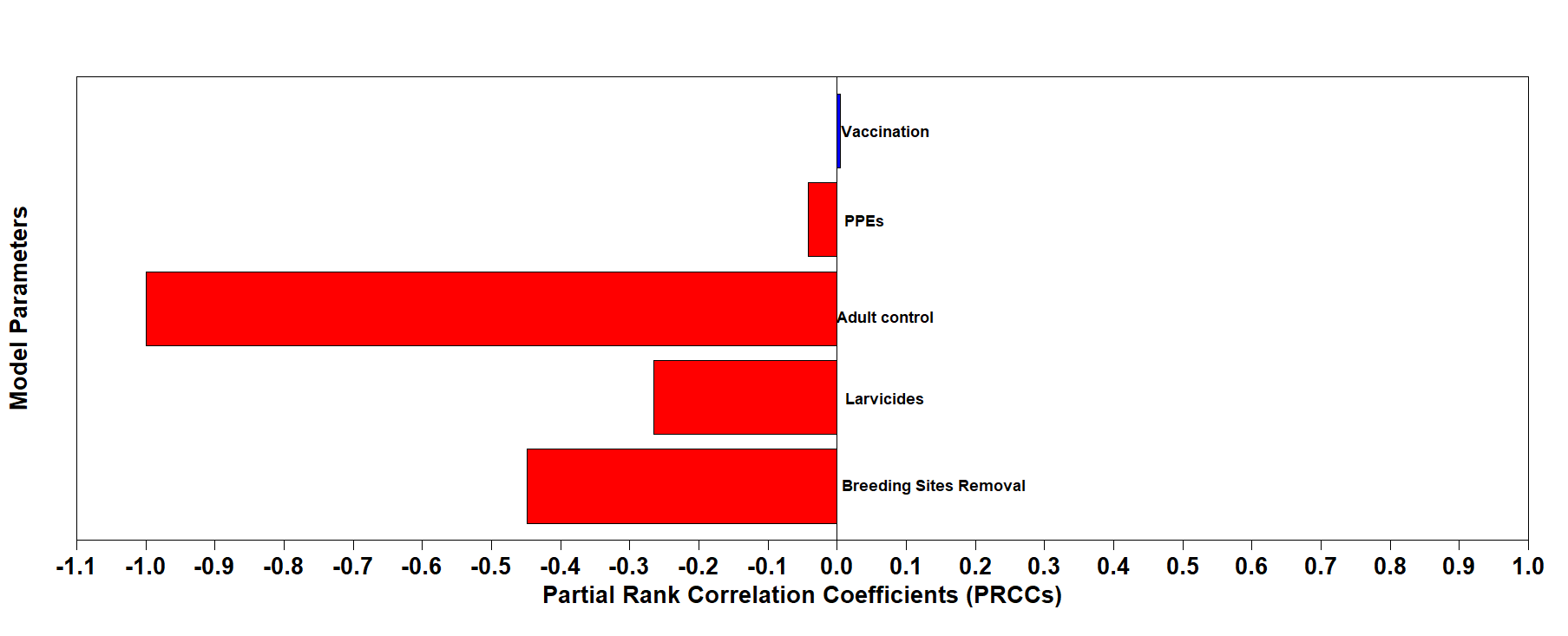}
			\caption{\small PRCC values for WNV-infected humans.} 
			\label{s1}
		\end{subfigure}
		\hfill
		\begin{subfigure}{0.9\textwidth}
			\hfill
			\includegraphics[width=\textwidth]{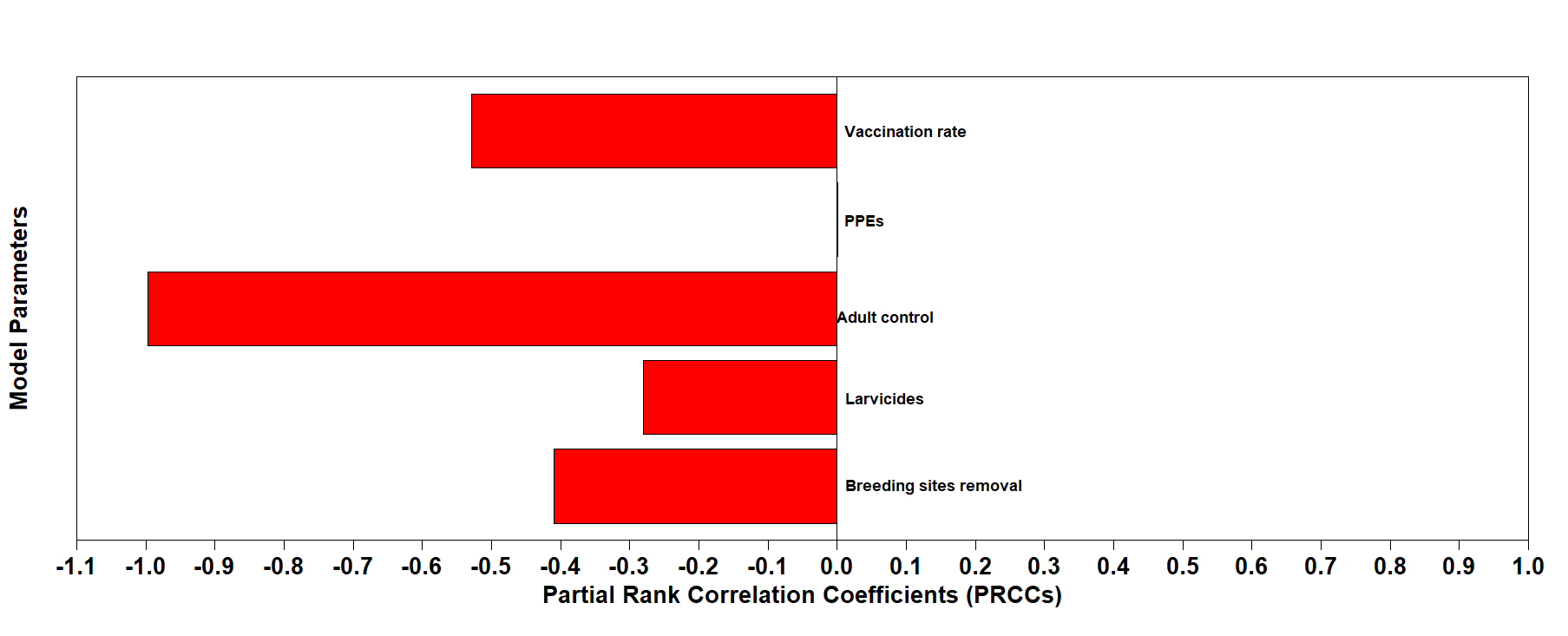}
			\caption{\small PRCC values for WNV-infected equids.} 
			\label{s2}
		\end{subfigure}
		\caption{\small Tornado plots showing the PRCC values for the control variables of the population dynamics of WNV infectious birds, WNV infected humans, and WNV infected equids.}
		\label{sens}
	\end{figure}

	\subsection{Behavioural dynamics of PPE adoption}\label{o1}
\begin{figure}[H]
		\centering
		\includegraphics[width=\textwidth]{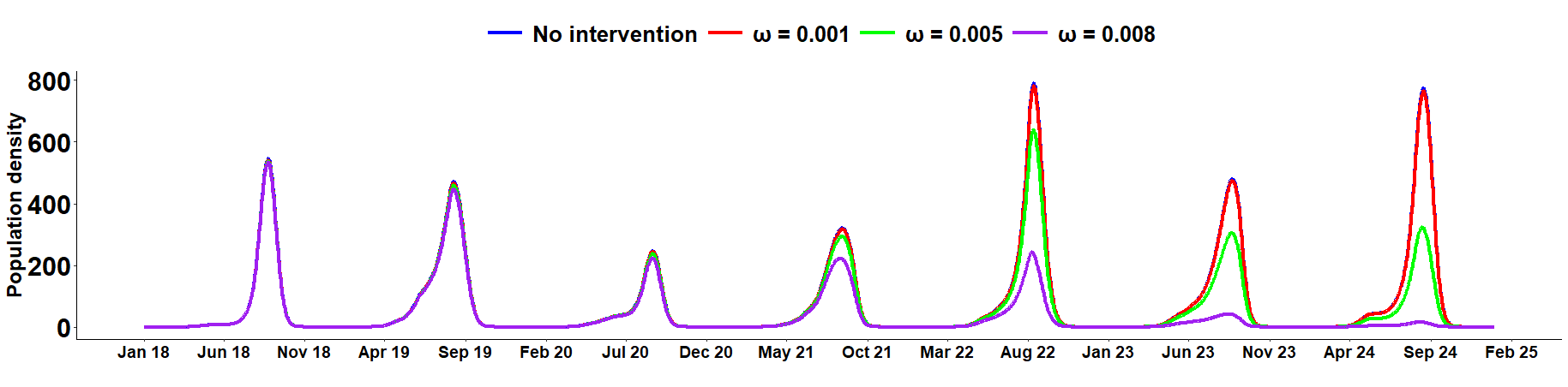}
		\caption{\small Population of infected humans for different values of the parameter $\omega$, representing the sensitivity of human behavioural response to WNV prevalence in birds.}
		\label{f4}
	\end{figure}

	Human adoption of PPE was examined in relation to WNV prevalence in birds (Fig. \ref{f4}). The parameter $\omega$, which represents the sensitivity of human behavioural response to increases in the infectious bird population, was varied such that PPE adoption begins to increase when $\dis I_B>\frac{1}{\omega}$. Simulations were performed for $\dis \kappa=4\times 10^{-4}$ and $\omega=0.001,\;0.005,\;0.008$. These values correspond to behavioural response thresholds of approximately 1000, 200, and 125 infectious birds, respectively. Smaller values of $\omega$ therefore represent a delayed behavioural response, in which PPE adoption occurs only when the outbreak becomes relatively large, whereas larger values represent earlier adoption of PPE as the infectious bird population begins to rise.
	
	\subsection{Impact of vaccinating equids}
	
\begin{figure}[H]
		\centering
		\includegraphics[width=\textwidth]{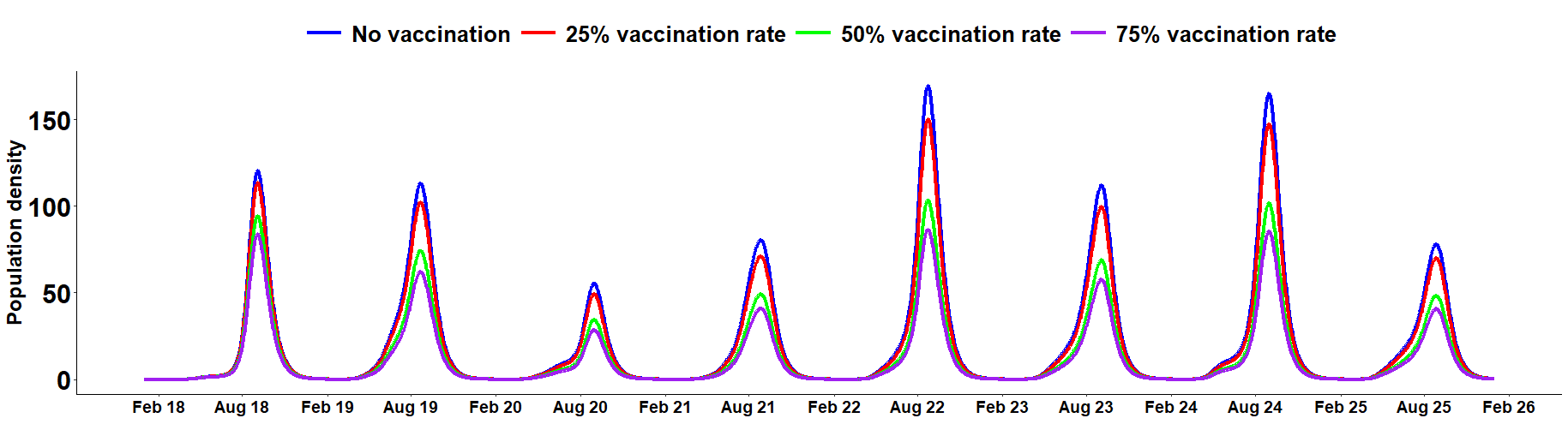}
		\caption{\small Population dynamics of infected under different vaccination rates. The blue curve represents the scenario without vaccination. The red, green, and purple curves correspond to annual vaccination rates of 25\%, 50\%, and 75\%, implemented as daily rates of 0.25/365, 0.5/365, and 0.75/365, respectively.}
		\label{fc4}
	\end{figure}

	The effect of the vaccination rate on WNV control was assessed by varying the vaccination rate between 0 and 1 (Fig. \ref{fc4}).

\subsection{Optimal control problem}\label{opt}
	
The control functions for the optimal control problem were derived analytically using Pontryagin's Maximum Principle, together with the characterisation and adjoint system (Section S4 of the Supplementary material). Simulations (Fig. \ref{bs}) indicate that intervention efforts are concentrated early in the transmission season. Breeding site removal and larviciding and adult control reach their maximum peak values on 11 May, 10 May and 24 April of the first year of simulation (Fig. \ref{bs}), respectively. In contrast, the peak number of infectious birds occurs later on 13 August. Thus, the maximum effect of control by breeding-site removal occurs approximately 13 weeks before the peak in infectious birds. Given that the objective was to minimise mosquito populations and infectious birds, while penalising adulticiding, the average control intensities remained relatively low throughout the season (with $\bar{u}_1=0.0031, \;\bar{u}_2=0.0024,\; \bar{u}_3=0.00059),$ highlighting that the optimal control strategy relies mainly on the first two intervention measures, while the adult control is applied only at lower levels.
	
	\begin{eqnarray}\label{fg3}
		\begin{cases}
			u_{1}^{\star}(t)=\min\left\{0.5,\max\left\{\dis 0,\frac{(b_L\delta_MN_M+L_M)(1-L_M/K_M)\lambda_{1}}{c_1}\right\}\right\},\\[10pt]
			u_{2}^{\star}(t)=\min\left\{0.5,\max\left\{\dis 0,\frac{L_M(1-L_M/K_M)\lambda_1}{c_2}\right\}\right\},\\[10pt]
			u_{3}^{\star}(t)=\min\left\{0.5,\max\left\{\dis 0,\frac{S_M\lambda_2+E_M\lambda_3+I_M\lambda_4}{c_3}\right\}\right\},\\[10pt]
		\end{cases}
	\end{eqnarray}

	\begin{figure}[H]
		\centering
		\begin{subfigure}{0.8\textwidth}
			\centering
			\includegraphics[width=\textwidth]{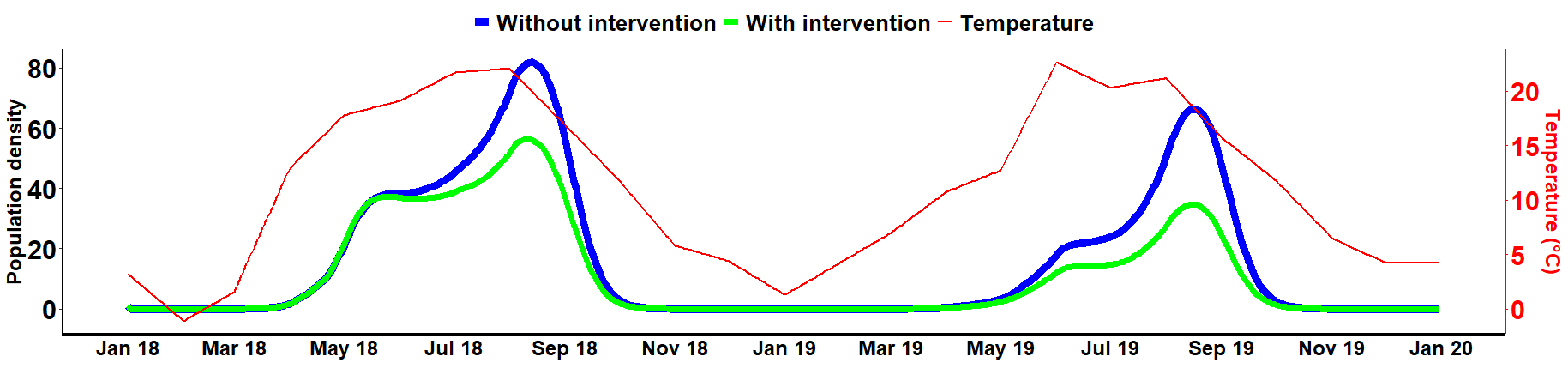}
			\caption{Infectious birds with and without interventions.}
			\label{b0}
		\end{subfigure}
		\hfill
		\begin{subfigure}{0.8\textwidth}
			\centering
			\includegraphics[width=\textwidth]{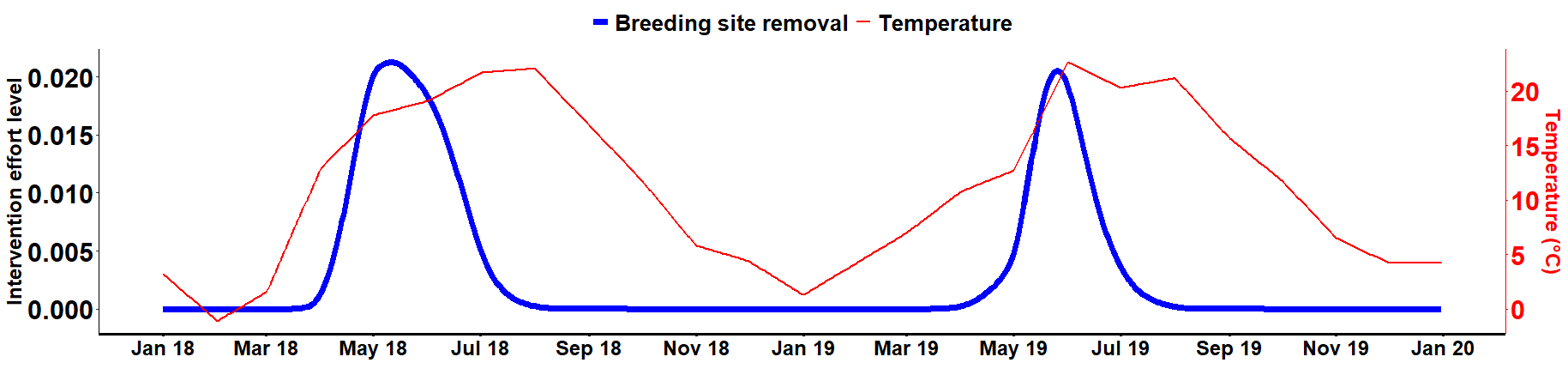}
			\caption{Physical removal and destruction of potential mosquito breeding sites $(u_1)$.}
			\label{b1}
		\end{subfigure}
		\hfill
		\begin{subfigure}{0.8\textwidth}
			\centering
			\includegraphics[width=\textwidth]{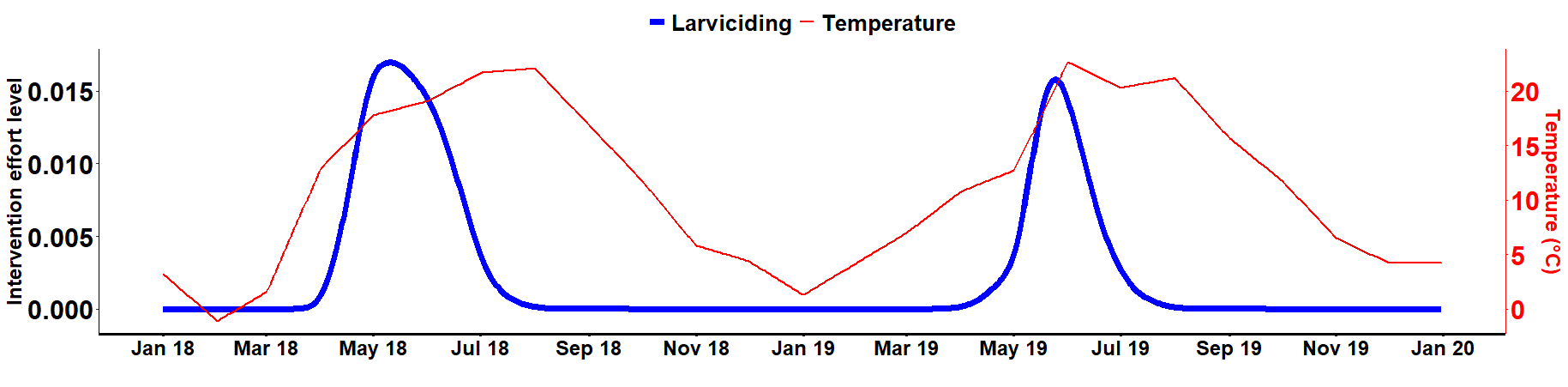}
			\caption{Larviciding $(u_2).$}
			\label{b2}
		\end{subfigure}
		\hfill
		\begin{subfigure}{0.8\textwidth}
			\centering
			\includegraphics[width=\textwidth]{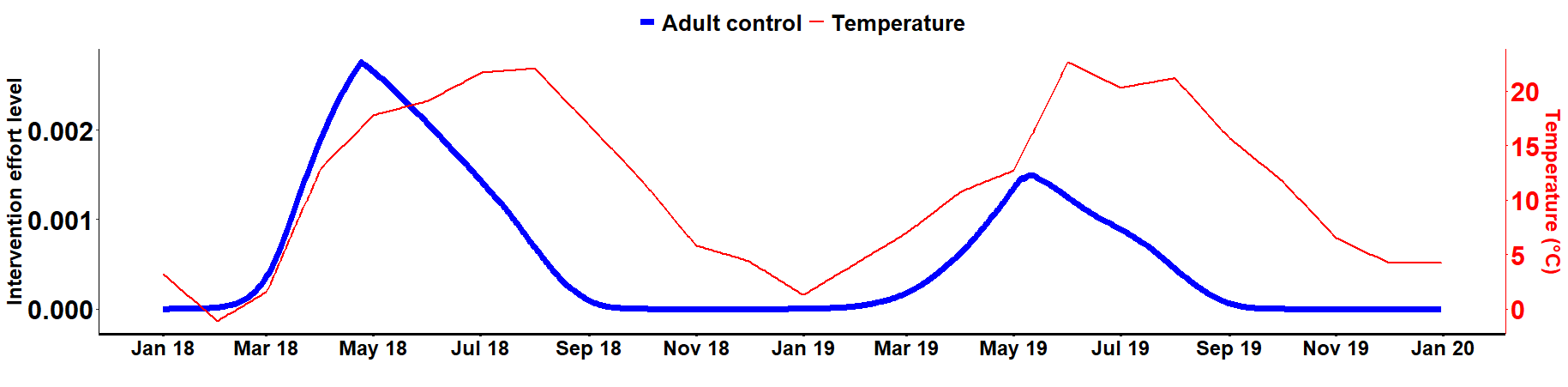}
			\caption{Adult control $(u_3).$} 
			\label{b3}
		\end{subfigure}
		\caption{Simulation of the infectious bird population compared to the control profile.}
		\label{bs}
	\end{figure}
	
\section{Discussion and conclusions}
	
We investigated the impact of different control measures against WNV. We adapted the model originally developed by \cite{rubel2008} to describe the dynamics of Usutu virus in Austria and later extended by \cite{Laperriere2011} to capture the transmission dynamics of WNV among birds, humans, and equids. Recent advances in interdisciplinary research have enabled the incorporation of human behavioural responses into epidemiological models. In particular, we modelled the adoption of PPE using imitation dynamics derived from evolutionary game theory \citep{Bauch2005}. This framework allows the integration of mosquito ecology, epidemiological dynamics, and human behavioural responses, while accounting for factors such as perceived inconvenience, limited awareness, or misinformation that may reduce the willingness to adopt protective measures.
	
Using temperature data for Berlin, we simulated the population dynamics of infectious birds, dead birds and infected equids, for the period 2018-2025 (Fig. \ref{vc}). Our model accurately predicted the known seasonal pattern for WNV dynamics in temperate regions. Some years had more cases than others, highlighting the impact of seasonal variation in yearly temperatures. To assess the model's performance, monthly averaged predictions were compared with observed surveillance data over the study period (Fig. \ref{v}). The model successfully reproduced the seasonal dynamics of WNV transmission, capturing the overall seasonal timing and monthly peaks of occurrences. Particularly in 2020 and 2022-2025, the model correctly identified peak months for both infectious and dead birds, yielding average Spearman correlation coefficients of 0.53 and 0.52, respectively, between observed and predicted seasonal patterns. For equids, the model identified the July peak in 2020 and the August peaks in 2024 and 2025, yielding $\rho=0.63$ for the overall WNV dynamics. Although agreements are evident in some years, discrepancies between observed and predicted patterns may partly be attributed to the lack of systematic WNV surveillance in Germany.
	
We explored different levels of individuals' reactions to disease prevalence in birds $(\omega)$ and how these influence PPE adoption and WNV infections in humans (Fig. \ref{f4}). The simulations show that when individuals adopt PPE promptly following reports of WNV cases in birds, the number of human infections can be substantially reduced (Section \ref{o1}). In particular, when individuals respond quickly to increases in WNV prevalence in birds (larger $\omega$), PPE adoption occurs earlier and reduces spillover transmission to humans. This result is consistent with several studies indicating that WNV infections in mosquitoes and amplifying hosts are typically detected at least two weeks before human cases are reported \citep{Angelini2010, Petrovi2018, Veksler2009}. Conversely, delayed responses (smaller $\omega$) prolong periods of unprotected human exposure to infectious mosquitoes before protective behaviour increases, leading to higher numbers of human infections. This result further highlights the importance of early risk communication based on bird surveillance, which can encourage the timely adoption of PPE and reduce human infections. 

Simulations of equid vaccination indicate that increasing the vaccination rate leads to a substantial reduction in WNV infections among equids. As shown in Fig. (\ref{fc4}), higher vaccination rate results in a decline in the number of infected equids. However, even at relatively high vaccination rates, the vaccine alone may not eliminate equid infections completely as some vaccines mainly reduce the number of viremic horses, duration and severity of clinical signs and WNV-mortality \citep{Bowen2014}. Indeed, when vaccination is complemented with mosquito-reduction strategies (various combinations in the shiny app), the model predicts a much larger decline in equid infections, a result supported by \cite{Cendejas2024}.

Sensitivity analysis of the seasonal recruitment function revealed that the simulated epidemic dynamics were reproduced more accurately for values of $p$ close to one (Fig. \ref{wb}). This suggests that recruitment functions with a temporal timing and magnitudes similar to those observed for northern goshawks may produce epidemic patterns that more closely match the observed dynamics. Importantly, this result does not imply that northern goshawks dominate the avian community, but instead emphasises how recruitment timing and magnitude shape the availability of susceptible hosts.

Results from the sensitivity analysis of the control parameters (Fig. \ref{sens}) indicate that the most effective control method is the use of adult controls, as it yields the highest PRCC values across all hosts. This result is consistent with the conclusion of \cite{BOWMAN2005}, who stated that, on average, adult control strategies, such as adulticiding, are more effective at preventing WNV spread in humans than personal protection. Breeding site removal follows on the list. Larviciding and PPE are also negatively correlated with the infected human population, highlighting that larviciding with perfect seasonal timing is highly effective, as also indicated by \cite{abdelrazec2015dynamics}. The adoption of PPE by humans reduces the population of infected humans, further complementing mosquito control efforts. On the other hand, vaccination of equids is also highly negatively correlated with the population of infected equids, indicating the need to intensify vaccination campaigns and efforts to curb WNV infections in equids.

Temperature fluctuations may also have economic implications, as larger outbreaks require increased resources for control, whereas smaller outbreaks require comparatively fewer resources \citep{Barber2010}. For this reason, an optimal control problem that considers temperature- time-dependent controls was studied. Results obtained suggest that breeding site removal and larviciding started before 1 April, aligning with rising spring temperatures, and maximised around 11 May, may have the potential to reduce the August peak in bird infections. Because the optimal controls reach their highest levels several weeks before the peak in infectious birds, the model results indicate that preventive measures should be implemented well before the epidemic peak, when mosquito populations begin to increase. This aligns with the suggestions from the \citet{Ecdc2023}, of preventively applying mosquito control methods before and during the transmission season, while starting awareness campaigns and other protective measures at the start of the WNV transmission season. On the other hand, adult control methods such as modifying overwintering sites might be implemented earlier (before March). While this result is theoretical, it is supported by other modelling studies, including \cite{Fesce2025} and \cite{Thomas2009}. However, adulticiding in winter, for example, is not an established control method and raises several questions about its practicality (e.g., indoor use of insecticides) that require further research. Additionally, this analysis indicates that temperature monitoring and forecasts can act as early warning indicators for the timely deployment of control measures.

Finally, a Shiny application was developed to explore the potential impact of future climate projections on WNV transmission and control. The app enables users to evaluate various intervention strategies under different climate scenarios without requiring mathematical/computational skills, thereby supporting climate-informed surveillance and control planning. Users can define the start and end of both the simulation and the control implementations. From the app, users can explore the effects of different control combinations at different time periods. Explorations indicate that although interventions reduce early-season transmission, they may also delay the epidemic peak within the same period. This occurs because early control suppresses infections and preserves a larger susceptible population, which can enable a later resurgence once interventions end and environmental conditions remain favourable for transmission. Such rebound dynamics can be seen in the app, and they are mainly because, in uncontrolled scenarios, high infection levels naturally decrease the number of susceptible hosts, whereas control measures reduce infections and preserve a larger susceptible population. A similar pattern occurs in host species alone (birds, humans and equids), given that infected birds often die from WNV, whereas humans and equids typically recover with prolonged immunity \citep{Papa2011, Tolnai2025}.
	
Mosquito elimination methods alone may not fully reduce the risk of arbovirus transmission, but they remain important for suppressing WNV risk across different host groups. To complement them, an integrated approach that combines mosquito control methods, human PPE adoption, and equid vaccination is the best strategy to reduce the WNV burden in humans and equids. For bird infections, more research is still needed to determine which methods can complement mosquito reduction efforts to lower infection rates. A study by \cite{Malik2018} provided a theoretical framework that considers bird vaccination and other methods to reduce contact between mosquitoes and birds, as birds are the amplifying hosts of WNV. This suggests that mosquito reduction efforts remain an important strategy for disrupting the WNV transmission cycle. 

Our WNV control model can be further improved in the future, e.g., by considering other mosquito control efforts, such as the use of modified mosquitoes, including sterile insect techniques and \textit{Wolbachia} mosquitoes \citep{Atyame2015, Lees2015}. Furthermore, other environmental factors, such as precipitation, wind speed, and direction, can be incorporated to support temperature in long term predictions under climate change \citep{Bhowmick2025, Mbaoma2024}. In recent years, advanced epidemiological modelling approaches, including system identification techniques and artificial intelligence-based adaptive policy frameworks, have been developed to improve the prediction and control of epidemic outbreaks \citep{Farooq2022, Tutsoy2021}. These approaches support the growing role of data-driven methods in complementing traditional mechanistic epidemic models, and this forms a future direction for the extension of this work. 

\section{Limitations of the study}

The lack of a system surveillance for WNV in Germany limits the reliability of observed cases. For model fitting, we relied on the WNV data from \cite{klji}, which are spatially and temporally sparse and subject to under-reporting, particularly regarding the large proportion of asymptomatic cases, resulting in incomplete detection of local WNV activity. In a real-world situation, uncertainty is a natural part of epidemiological modelling. It can range from incomplete knowledge of model parameters (parametric uncertainty), assumptions made in the model structure, and external factors like climate variability. Both the size and form of these uncertainties is often unknown, and hence, they must be taken into account when interpreting model predictions \citep{H2009, King2015, Li2024}. Moreover, our WNV control strategies are scenario-based. Thus, due to the absence of field-tested intervention data in Germany, the study aims to provide a qualitative assessment of some of the available WNV control strategies.

\section*{CRediT authorship contribution statement}
\textbf{Pride Duve:} Designed the study, developed the model, proved the mathematical properties, computed the numerical simulations, and wrote the original draft. \textbf{Felix Gregor Sauer:} Supervised the study, designed the model, and reviewed the final draft. \textbf{Renke Lühken:} Supervised the study, designed the model, and reviewed the final draft. 
\section*{Declaration of competing interests}
The authors declare that they have no known competing financial interests or personal relationships that could have appeared to influence the work reported in this paper.
\section*{Data availability}
Original datasets can be obtained from the \citet{klji} website.
\section*{Acknowledgements}
The authors would like to thank the Federal Ministry of Education and Research of Germany (BMBF) under the project NEED (Grant Number 01Kl2022), the Federal Ministry for the Environment, Nature Conservation, Nuclear Safety and Consumer Protection (Grant Number 3721484020), and the German Research Foundation (JO 1276/51) for funding this project. The authors would like to thank Markus Jansen for his support in running our app in a Docker container on the BNITM website.

\newpage

\begin{table}[H]
	\caption{\small Definition of parameters used in system (\ref{sys1a}).}
	\renewcommand{\arraystretch}{1.8}
	\small
	\centering
	\begin{adjustbox}{width=1.05\textwidth}
		\begin{tabular}{cccc}
			\toprule 
			\textbf{Parameter} & \textbf{Definition}&\textbf{Value}& \textbf{Source} \\
			\midrule 		
			$\delta_M(D)$ &fraction of non-overwintering mosquitoes&Eq. (\ref{dM})&\citep{rubel2008}\\
			$b_L(T)$ &birth rate of mosquito larva&Eq. (\ref{eq1})&\citep{rubel2008}\\
			$\mu_L(T)$ &mortality rate of mosquito larva&Eq. (\ref{eq3})&\citep{rubel2008}\\
			$b_M(T)$ &birth rate of adult mosquitoes&Eq. (\ref{eq4})&\citep{rubel2008}\\
			$k(T)$ &mosquito biting rate&Eq. (\ref{eq5})&\citep{rubel2008}\\
			$\lambda_{BM}(T)$ &force of infection on mosquitoes&Eq. (\ref{eq6})&\citep{rubel2008}\\
			$\mu_M(T)$ &mortality rate of adult mosquitoes&Eq. (\ref{eq7})&\citep{rubel2008}\\
			$\gamma_M(T)$ &latency rate of mosquitoes&Eq. (\ref{eq8})&\citep{rubel2008}\\
			$p_{M}$ &transition probability by infectious mosquitoes&$1.0$&\citep{Laperriere2011}\\
			$K_{M}$ &carrying capacity of mosquito population&3,300,000&\citep{Laperriere2011}\\
			NVmin&minimum number of adult mosquitoes&500,000&\citep{Laperriere2011}\\
			$b_{B}$ &crude per-capita daily recruitment rate of susceptible birds&Eq. (\ref{h})&defined by the authors\\
			$\mu_{B}$ & crude per-capita natural mortality rate of birds&$0.0005$&\citep{Mbaoma2024}\\
			$\gamma_{B}$ & latency rate of birds&$0.196$&\citep{Mbaoma2024}\\
			$\lambda_{MB}(T)$ &force of infection on birds&Eq. (\ref{eq10})&\citep{rubel2008}\\
			$\phi_{B}$ & mosquito to bird ratio&$50-300$&defined by the authors\\ 
			$\nu_{B}$ & proportion of infectious birds that die due to WNV&$0.655$&\citep{Mbaoma2024}\\ 
			$\alpha_{B}$ & removal rate of infectious birds&$0.867$&\citep{Mbaoma2024}\\ 
			$p_{B}$ &transition probability by infectious birds&$0.125$&\citep{Laperriere2011}\\ 
			$K_{B}$ &carrying capacity of bird population&1,159,340&defined by the authors\\
			$b_{H}$ &recruitment rate of susceptible humans&0.0000452&defined by the authors\\
			$\mu_{H}$ & natural mortality rate of humans&0.0000287&defined by the authors\\
			$\gamma_{H}$ & latency rate of humans&0.25&\citep{Laperriere2011}\\
			$\lambda_{MH}(T)$ &force of infection on humans&Eq. (\ref{eq13})&\citep{Laperriere2011}\\
			$\phi_{H}$ & mosquito to human ratio&0.3-3&defined by the authors\\ 
			$\nu_{H}$ & proportion of infected humans that die due to WNV&0.004&\citep{Laperriere2011}\\ 
			$\alpha_{H}$ & removal rate of infected humans&0.5&\citep{Laperriere2011}\\ 
			$\kappa$ &imitation rate&0.00001&defined by the authors\\
			$\omega$ & individuals' sensitivity to the WNV prevalence in birds&varies&defined by the authors\\
			$b_{E}$ &birth rate of susceptible equids&0.00016&\citep{Laperriere2011}\\
			$\mu_{E}$ & natural mortality rate of equids&0.00011&\citep{Laperriere2011}\\
			$\gamma_{E}$ & latency rate of equids&0.05&\citep{Laperriere2011}\\
			$\lambda_{ME}(T)$ &force of infection on equids&Eq. (\ref{eq14})&\citep{Laperriere2011}\\
			$\phi_{E}$ & mosquito to equid ratio&3&defined by the authors\\ 
			$\nu_{E}$ & proportion of infected equids that die due to WNV&0.04&\citep{Laperriere2011}\\ 
			$\alpha_{E}$ & removal rate of infected equids&0.2&\citep{Laperriere2011}\\ 
			$\theta_{E}$ & vaccination rate of equids&[0-1]&defined by the authors\\ 
			$\omega_{E}$ & vaccine waning rate&0.0019&defined by the authors\\ 
			$c_{E}$ & vaccine effectiveness&0.5&defined by the authors\\ 
			$u_1$ & removal and destruction of potential mosquito breeding sites&varies&defined by the authors\\  
			$u_2$ & larvicides&varies&defined by the authors\\
			$u_3$ & adult control&varies&defined by the authors\\
			\bottomrule
		\end{tabular}
	\end{adjustbox}
	\label{t2}
\end{table}

\bibliographystyle{plainnat}  

\bibliography{References}      

\end{document}